\documentstyle[psfig]{cupconf}

\ifoldfss
\else
  \ifnfssone
    \newmathalphabet{\mathit}
      \addtoversion{normal}{\mathit}{cmr}{m}{it}
      \addtoversion{bold}{\mathit}{cmr}{bx}{it}
    \newmathalphabet{\mathcal}
      \addtoversion{normal}{\mathcal}{cmsy}{m}{n}
    \else
    \ifnfsstwo
    \fi
  \fi
\fi

\def\lsim{~\rlap{$<$}{\lower 1.0ex\hbox{$\sim$}}}
\def\gsim{~\rlap{$>$}{\lower 1.0ex\hbox{$\sim$}}}
\def\Msun {M_{\odot}\ }

\title[Stellar Populations in Dwarf Galaxies]{Stellar 
Populations in Dwarf Galaxies: A Review of the Contribution of 
HST to our Understanding of the Nearby Universe}

\author[Tolstoy]%
{Eline Tolstoy\thanks{Present address:
UK Gemini Support Group, University of Oxford, Keble Rd, 
Oxford, UK 
(email: etolstoy@astro.ox.ac.uk)
}}

\affiliation{European Southern Observatory, Karl-Schwarzschild str. 2,
Garching bei M\"{u}nchen, Germany}

\begin{document}
\ifnfssone
\else
  \ifnfsstwo
  \else
    \ifoldfss
      \let\mathcal\cal
      \let\mathrm\rm
      \let\mathsf\sf
    \fi
  \fi
\fi

\maketitle

\begin{abstract}

This review aims to give an overview of the contribution of the {\it
Hubble Space Telescope} to our understanding of the detailed
properties of Local Group dwarf galaxies and their older stellar
populations.  The exquisite stable high spatial resolution combined
with photometric accuracy of images from the {\it Hubble Space
Telescope} have allowed us to probe further back into the history of
star formation of a large variety of different galaxy types with
widely differing star formation properties. It 
has allowed us to extend our studies out to the edges of
the Local Group and beyond with greater accuracy than ever before.  We
have learnt several important things about dwarf galaxy evolution from
these studies. Firstly we have found that no two galaxies have
identical star formation histories; some galaxies may superficially
look the same today, but they have invariably followed different
paths to this point. Now that we have managed to probe deep into the
star formation history of dwarf irregular galaxies in the Local Group
it is obvious that there are a number of similarities with the global
properties of dwarf elliptical/spheroidal type galaxies,
which were previously thought to be quite distinct. The
elliptical/spheroidals tend to have one or more discrete episodes of
star formation through-out their history and dwarf irregulars are
characterized by quasi-continuous star-formation. The previous strong
dichotomy between these two classes has been weakened by these new
results and may stem from the differences in the environment in which
these similar mass galaxies were born into or have inhabited for most
of their lives. The more detailed is our understanding of star
formation processes and their effect on galaxy evolution in the nearby
Universe the better we will understand the results from studies of the
integrated light of galaxies in the high-redshift Universe.

\end{abstract}


\section{Introduction}

This review is a survey of the {\it Hubble Space Telescope} (HST) 
studies of resolved stellar
populations of nearby galaxies that have determined accurate global
star formation histories going back several Gyr. The determination of
such a detailed the star formation history depends upon the ability to
accurately photometer many Gyr old stars.  This was possible from
ground based imaging for our nearest companions, the dwarf
spheroidals, and the Magellanic Clouds. HST, with it's outstanding
combination of lower sky brightness, high resolution, and a stable and
constant point spread function, has allowed us to to probe further out
into the Local Group and even beyond this to look at a much more
diverse sample of galaxies than previously possible, and it has also
allowed us renewed insights into our nearest neighbours.

I am leaving out any discussion of the considerable body of literature
on the HST observations of the stellar populations of the Magellanic
Clouds ({\it e.g.}, Holtzman {\it et al.} 1999; Panagia {\it et al.}
2000) which technically do not count as dwarf galaxies ({\it e.g.},
Tammann 1994; Binggeli 1994) but are often assumed to be so.  Because
these galaxies are so large on the sky it is difficult for HST by
itself to gain a perspective of the global star-formation properties,
although attempts are being made ({\it e.g.}, Smecker-Hane {\it et
al.} 1999).  There are also numerous HST studies of stellar
populations at the small scale of individual star clusters and H~II
regions in the Clouds ({\it e.g.}, Da~Costa~1999; Massey~1999;
Hunter~1999).  Studies of the stellar populations of the Magellanic
Clouds could easily take up an entire review, not to say a conference,
in their own right. I have also neglected considerable work on
resolved stellar populations of dwarfs in the UV ({\it e.g.} Brown
{\it et al.} 2000; Cole {\it et al.} 1998), because the detailed
connection to quantifying a star formation history is unclear. This is
also true of studies of star clusters around nearby dwarf galaxies
({\it e.g.}, Da~Costa~1999; Hodge {\it et al.} 1999; Mighell,
Sarajedini \& French 1998).

\begin{table} 
  \begin{center} 
  \caption{The Local Group}

  \begin{tabular}{llllccl} 

 name& type    & $M_V$   & $\Sigma_0$& M$_{tot}$& M$_{tot}$/M$_{HI}$ &[Fe/H] \\[3pt] 
			&	  &	    &mag arcsec$^{-2}$&10$^6\Msun$&& dex  \\
\\
{\it Spiral galaxies:}   &&&&&&\\
	M~31		& Sb      & $-$21.2 & 10.8&2$\times 10^6$& 0.002& +0.2	\\	
	Milky Way	& Sbc	  & $-$20.9: &     & 10$^6$& 0.004 & ~~~0.		\\	
	M~33		& Sc	  & $-$18.9 & 10.7&10$^5$&0.02 &$-$0.2 		\\	
\\
{\it Irregular galaxies:}&&&&&&\\
	NGC~3109	& Irr	  & $-$15.7 & 23.6$\pm0.2$&6550	& 0.11	&$-1.2\pm0.2^{a}$\\	
	LMC		& Irr	  & $-$18.1 & 20.7&6000	& 0.5	&$-$0.7		\\	
	SMC		& Irr	  & $-$16.2 & 22.1& 2000& 0.25	&$-$1.0		\\	
\\
{\it Dwarf Ellipticals:} &&&&&&\\
	M~32		& dE2	  & $-$16.7 & $<$11.5:&2120& $<$0.001&$-1.1\pm0.2$\\	
        NGC~205		& dE5	  & $-$16.6 & 20.4$\pm0.4$& 740	& 0.001	&$-0.8\pm0.1$\\	
	NGC~185		& dE3     & $-$15.5 & 20.1$\pm0.4$& 130	& 0.001	&$-1.2\pm0.15$\\	
	NGC~147 	& dE4	  & $-$15.5 & 21.6$\pm0.2$& 110	& $<$0.001&$-1.1\pm0.2$\\	
	Sagittarius	& dE7	  & $-$13.4 & 25.4$\pm0.3$& 500:& $<$0.0001&$-1.0\pm0.2$\\	
\\
{\it Dwarf Irregulars:}  &&&&&&\\
	NGC~6822	& dIrr	  & $-$15.2 & 21.4$\pm0.2$&1640 & 0.08	&$-0.7\pm0.2^{a}$\\
	IC~10 		& Irr	  & $-$15.7 & 22.1$\pm0.4$& 1580& 0.10	&$-0.7\pm0.15^{a}$\\	
	IC~1613		& dIrr	  & $-$14.7 & 22.8$\pm0.3$& 795 & 0.07	&$-1.1\pm0.2^{a}$\\	
	IC~5152		& dIrr	  & $-$14.8 &     & 400 & 0.15	&$-0.6\pm0.2^{a}$\\	
	Sextans~B	& dIrr	  & $-$14.2 &    & 885  & 0.05	&$-1.1\pm0.3^{a}$\\	
	Sextans~A	& dIrr	  & $-$14.6 & 23.5$\pm0.3$& 395 & 0.20	&$-1.4\pm0.2^{a}$\\	
       	WLM   		& dIrr    & $-$14.5 & 20.4$\pm0.05$& 150 & 0.40	&$-1.1\pm0.2^{a}$\\ 
	Phoenix		& dIrr/dE & $-$10.1 &     &  33 & 0.006:&$-1.9\pm0.1$\\	
	Pegasus		& dIrr	  & $-$12.9 &     &  58 & 0.09	&$-1.0\pm0.14^{a}$\\	
	LGS~3		& dIrr/dE & $-$10.5 & 24.7$\pm0.2$&  13 & 0.03	&$-1.8\pm0.3$\\	
	Leo~A		& dIrr	  & $-$11.4 &     &  11 & 0.72	&$-1.6\pm0.15^{a}$\\	
	SagDIG		& dIrr	  & $-$12.3 & 24.4$\pm0.3$&  9.6& 0.92	&$-1.5\pm0.3^{a}$\\	
	DDO~210		& dIrr	  & $-$10.0 & 23.0$\pm0.3$&  5.4& 0.35	&$-1.9\pm0.12$\\	
	EGB 0427+63	& dIrr	  & $-$12.6 & 23.9$\pm0.1$&     &	&$-1.4\pm0.1^{a}$\\
\\
{\it Dwarf Spheroidals:} &&&&&&\\
	Fornax		& dE3	  & $-$13.2 & 23.4$\pm0.3$&  68 & $<$0.001&$-1.3\pm0.2$	\\	
	Ursa~Minor	& dE5	  & $-$8.9  & 25.5$\pm0.5$&  23 & $<$0.002&$-2.2\pm0.1$	\\	
	Draco		& dE3	  & $-$8.8  & 25.3$\pm0.5$&  22 & $<$0.001&$-2.1\pm0.15$\\	
	Leo~I		& dE3	  & $-$11.9 & 22.4$\pm0.3$&  22 & $<$0.001&$-1.5\pm0.4$	\\	
	Sextans		& dE4	  & $-$9.5  & 26.2$\pm0.5$&  19 & $<$0.001&$-1.7\pm0.2$	\\	
	Carina		& dE4	  & $-$9.3  & 23.9$\pm0.4$&  13 & $<$0.001&$-2.0\pm0.2$	\\	
	Sculptor	& dE	  & $-$11.1 & 23.7$\pm0.4$&  6  & $<$0.004&$-1.8\pm0.1$	\\	
	Antlia		& dE3	  & $-$10.8 & 24.3$\pm0.2$&  12 & 0.08	  &$-1.8\pm0.25$\\	
	Tucana		& dE5	  & $-$9.6  & 25.1$\pm0.06$&     & 	  &$-1.7\pm0.15$\\	
	Cetus		& dE4	  & $-$10.1 & 25.1$\pm0.1$&     & 	  &$-1.9\pm0.2$	\\	
	Leo~II		& dE0	  & $-$9.6  & 24.0$\pm0.3$&  10 & $<$0.001&$-1.9\pm0.1$	\\	
	And~I		& dE0	  & $-$11.9 & 24.9$\pm0.01$&     & 	  &$-1.5\pm0.2$	\\	
	And~II		& dE3	  & $-$11.1 & 24.8$\pm0.05$&     & 	  &$-1.5\pm0.3$	\\	
	And~III		& dE6	  & $-$10.3 & 25.3$\pm0.05$&     &         &$-2.0\pm0.2$ \\	
	And~V		& dE3	  & $-$9.1  & 24.8$\pm0.2$&    &	  &$-1.6\pm0.2$	\\
	And~VI (Peg dSph)& dE3	  & $-$11.3 & 24.3$\pm0.05$&     &	  &$-1.6\pm0.2$	\\
	And~VII (Cass dSph)& dE3     & $-$12.0 & 23.5$\pm0.05$&     &	  &$-1.4\pm0.3$	\\

  \end{tabular}
$^a$ these values were converted from [O/H] measurements, assuming constant [Fe/O]=0.
  \end{center} 
\end{table} 

\subsection{Dwarf Galaxy Types:}

Classification is a difficult and often emotive buisness, and I do
not attempt to seriously address this issue, except to make it easier
for me to refer to the sample of galaxies in the Local Group in broad
terms, and with my particular interest in their star-forming
properties.  So, bearing this caveat in mind let me introduce the four
different classes of dwarf galaxies that should cover anything out
there:

{\it Dwarf Irregular} (dI) galaxies are arguably the most common type
of galaxy in the Universe ({\it cf}. Ellis 1997), they are not clustered
around larger galaxies, but appear to have a fairly random
distribution through out the Local Group, and indeed in the Universe.
They are usually loosely structured late-type gas rich systems with
varying levels of star formation occurring in a haphazard manner across
the galaxy.  The velocity field of the HI gas in these systems can be
dominated by random motions rather than rotation ({\it e.g.}, Lo, Sargent \&
Young 1993, but see Skillman 1996), for the fainter dIs ({\it e.g.}, Leo~A;
v$_{rot} \sim 5$km/s), but for the more massive dIs ({\it e.g.}, NGC~6822,
Sextans~A) solid body rotation is clearly seen, with amplitudes of
$30-40$km/s.

{\it Blue Compact Dwarf} (BCD) galaxies are gas rich systems dominated
by a region of extremely active star formation, and resembling the
massive HII regions which can be found in larger galaxies. They are
thought to be forming stars at a rate which they can only maintain for
a short period ({\it e.g.}, Searle, Sargent \& Bagnuolo 1973).  This type of
galaxy may be a dI undergoing a period of particularly active star
formation ({\it e.g.}, Tolstoy 1998a).  Within the Local Group, IC~10 is a
fairly good approximation of what we expect a BCD to look like (see
van den Bergh 2000; Hunter {\it et al.} in prep.), and perhaps also IC~5152.
The more distant BCDs could easily be embedded in larger
low surface brightness galaxies, which are easier to see 
within the Local Group. There are several, classical, examples of BCDs
just beyond the Local Group ({\it e.g.} NGC~1569 \& VII~Zw403).

{\it Dwarf Elliptical} (dE) galaxies are basically low luminosity
Elliptical galaxies, with smooth surface brightness distributions
({\it e.g.}, Ferguson \& Binggeli 1994).  They are typically dominated by an
old stellar population, but as Baade already noticed in 1951, they are
subject to the same extreme variations of stellar population as other
dwarf galaxy types. Baade (1951) found that the archetypal dEs NGC~185
and NGC~205 contain B stars along with gas and dust.  Recent detailed
analysis suggests that several epochs of star formation over long
time scales are needed to explain the characteristics of dE stellar
populations ({\it e.g.}, Ferguson \& Binggeli 1994; Han {\it et al.}
1997). Most of the bright dEs (M$_B < -16$) appear to have nuclei, and
there is some evidence that these are dynamically separate
super-massive star clusters ({\it e.g.}, M~54 in Sagittarius, Ibata et
al. 1994; and NGC~205, Carter \& Sadler 1990).  dEs are strongly
clustered with the largest galaxies, and four of the five dEs in the
Local Group are found in proximity to M~31.

{\it Dwarf Spheroidal} (dSph) galaxies are basically low-surface
brightness, non-nucleated dEs. Many argue that the dSph are merely the
low luminosity tail of the dE galaxy class ({\it e.g.}, Ferguson \& Binggeli
1994), and the fact that they are {\it often} clustered around bigger
galaxies such as our Galaxy, M~31, and perhaps also NGC~3109 tends to
support this. However it is also possible that at least a fraction of
this class fit into a common evolutionary scenario with dIs and BCDs,
their past star-formation having been dominated by bursts
({\it e.g.} Carina).  There are a number of so called {\it transition}
objects, such as Phoenix, Antlia and LGS~3 which are hard to fit
unambiguously into either the dE or dI class. They have very little or
no present-day star formation, and yet they have associated HI gas, so
it seems a fair assumption that they will form stars again before too
long, and will then clearly belong to the dI class. 

\subsection{The Properties of Local Group Galaxies}

Our local neighbourhood, the Local Group, is arguably as
representative a piece of the Universe as any ({\it e.g.}, van den Bergh
2000).  What we learn about the properties of star formation and
galaxy evolution here can justifiably be extrapolated to explain what
we see in the distant, early Universe.  That which is on our doorstep
provides us with the chance to properly understand the dominant
physical processes in great detail.

As with galaxy classification, there are different
selection criteria which vary the number of dwarf and irregular
galaxies included in a census of Local Group members.  I have
chosen to follow dynamical arguments ({\it e.g.}, Irwin 1998), which tends
to allow a slightly more distant limit to the distance from the centre
than van den Bergh (2000). There is no straight forward boundary
between the Local Group and the Sculptor, and many of the galaxies in
this region could ``belong to either'', by dynamical arguments.
Listed in Table~1 are some basic properties of the galaxies I have
assumed to be members of the Local Group.  For each galaxy there is
listed the absolute V magnitude (M$_V$), the central surface
brightness in V ($\Sigma_0$), the total mass (M$_{tot}$), and the
fraction of this total in HI gas, when I could find them.  
Also listed is an estimate of the
mean metallicity of each galaxy, given as [Fe/H], for the sake of
uniformity.  Where the metallicity was ``converted'' from [O/H], this
is noted. The rest of the values are predominantly based upon the
colour of the red giant branch, and thus ought to be treated as lower
limits. A lot of detail is glossed over in presenting a uniform
``metallicity'' for a range of galaxy type, as here ({\it cf}, Skillman 1998
for much more detail on this complex topic).

Most of the data in Table~1 came from Mateo (1998), which contains a
very complete collection of what is known about Local Group dwarf
galaxies.  Mateo also provides explanations of the error bars, and
where these data originally come from.  The purpose of Table~1 is largely
illustrative and I would recommend anyone who would like to use the
values in this table to look them up in Mateo, and even better
still in the original references provided there.  I also recommend the
detailed descriptions of the individual galaxies in van den Bergh
(2000).  The data for the more massive galaxies in the Local Group:
the Milky Way, M31, M33 and the LMC/SMC are extracted from various
standard sources, and not meant to be definitive, merely to illustrate
the scale sizes between dwarf galaxies and their large neighbours in
the Local Group. 

The mass of the Local Group is dominated by three large spiral
galaxies, namely our Galaxy, M~31 and M~33. However, the largest
population {\it by number} are the dwarf type galaxies (see Table~1).
The Local Group contains several examples of each
class of dwarf galaxy, as defined above. All dwarf galaxy types
could plausibly come from the same type of progenitor but for
reasons of differences in initial dark matter content, or environment
or chance encounters with other galaxies follow different evolutionary
paths which result in the different present-day properties
({\it e.g.}, Ferrara \& Tolstoy 2000; Binggeli 1994; Davies \& Phillips 1988)

\begin{figure} 
\vskip 1cm
\psfig{figure=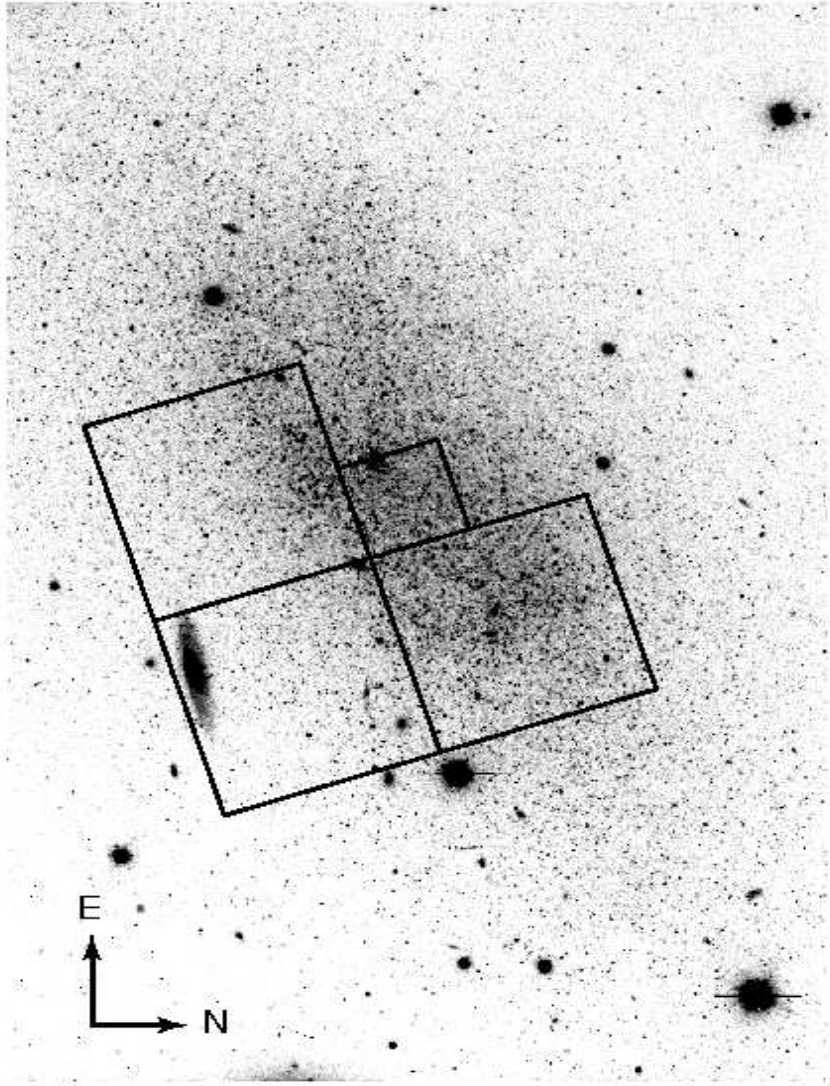,height=15cm,width=12cm}
\caption{
This is a WIYN 3.5m telescope image of the Pegasus dwarf taken in
0$^{\prime\prime}_.$6 seeing, from Gallagher {\it et al.} (1998). This is a
good example of a low surface brightness dI in the Local Group. It
shows the bright core where star formation is occurring, and the more
extended dE-like main body of the galaxy.  The field of view shown here is
4$^{\prime}_.8\times6^{\prime}_.$7 on a side, 
which corresponds to 1$\times$1.5~kpc at the distance of
Pegasus (760~kpc).  The location of the WFPC2 pointing is also shown.
}
\end{figure} 

The total mass of a galaxy has a critical impact on the ability of that
galaxy to form stars. It determines how effectively gas can be
compressed to increase the efficiency of star formation and also how
hard it is for supernova explosions to disrupt or even rid the system
of gas delaying or preventing future star formation (e.g, Mac Low \&
Ferrara 1999; De Young \& Heckman 1994; Dekel \& Silk 1986).  The
lowest mass dwarf galaxies are often at the hairy edge of being able
to retain their gas whilst forming stars, and many clearly have lost
the battle. For the lowest mass galaxies any small perturbation can
have a dramatic impact on the evolution of these systems, and it is
this sensitivity to initial conditions and random events which may 
explain some of the dispersion of the values in Table~1 (see 
Ferrara \& Tolstoy 2000).

One of the few obvious correlations between dwarf galaxy properties is
that between absolute magnitude (M$_V$) and global metallicity ({\it
e.g.}, Skillman, Kennicutt \& Hodge 1989).  If a reasonably uniform
mass to light (M/L) ratio can be assumed, then this can be interpreted
as a mass-metallicity relation. Those galaxies which fall
significantly off the relation ({\it e.g.} extreme BCDs) can be
assumed, with very good reasons, to have a significantly different
M/L. Indeed some scatter in the M/L of average galaxies might well
account for the scatter seen in the mass-metallicity relation ({\it
e.g.}, Ferrara \& Tolstoy 2000).  A large uncertainty in proving or
disproving this is the lack of reliable measurements of total (or
dynamical) masses of dwarf galaxies.

Basically, the global star-formation properties of a galaxy appear to
be dominated by the total mass.  This may perhaps vary when galaxies
interact and merge, but it is difficult to disentangle the effects of
merging two small galaxies to form a bigger one and the temporary
boost to the star formation rate from the merger.  These two effects
might balance each other - because taken at face value the
luminosity/mass-metallicity relation seems to say that a galaxy knows
how big it is from the earliest times. If a lot of dwarfs were added
together they would retain the global metallicity of the original
pieces, whilst increasing in luminosity. This simple arithmetic could
also argue against merging having a significant impact upon Local
Group galaxies, at least in recent times.

\section{Why Study Dwarf Galaxies?}

As demonstrated in Table~1 Local Group dwarf galaxies have a wide
range of different properties.  They span a large {\it mean}
metallicity range, down the lowest seen anywhere.  They also exhibit a
range of gas fractions, and density, from no gas all the way to gas
dominated.  They are also to be seen in a range of proximities to
other systems of varying mass. Thus a study of the dwarf galaxy
members of the Local Group allows us to study star formation over a
large range of initial conditions. When the star formation properties
of the Local group galaxies are looked at together ({\it e.g.}, Mateo
1998; Da Costa 1998; Grebel 1998), the only global statement that can
be made with no fear of contradiction is that no two are exactly
alike.

The smaller dwarf galaxies effectively have a single-cell mode of
global star formation, which in principle ought to be more straight
forward to comprehend than larger galaxies where different star
formation regions can apparently be unaware of each other or
interfere strongly with each other, or anything between these
extremes. This, including spiral density
waves, bars, jets and other dynamical effects can create severe
complications to the straight forward interpretation of the
relationship between gas and stars and star formation. Although, 
as can be seen by the results presented in this review, 
even these small galaxies are capable of a high degree of complexity,
so ``straight forward'' is a very relative statement.

Dwarf galaxies have the added benefit for HST observations, that they
are small enough, at the modest distances typical for Local Group members
that a significant fraction of the surface area of a dwarf 
can typically fit into the WFPC2 field of view (see Figure~1).

The Local Group contains galaxies whose star formation histories
should be typical of galaxy group members, and thus of star formation
throughout the Universe. It must therefore include remnants of the
epoch $\sim5-8$~Gyr before the present when actively star forming
galaxies produced the faint blue galaxy population seen at
intermediate redshifts.  We can directly measure star formation
histories of nearby galaxies back to the era of faint blue galaxies
with sufficiently deep and accurate imaging and using established
quantitative techniques for analysing colour-magnitude diagrams ({\it
e.g.}, Tosi {\it et al.} 1991; Tolstoy 1996; Aparicio {\it et al.}
1996).

Consistent with the properties of faint blue galaxies in redshift
surveys, dwarf galaxies appear to have erratic star formation rates,
and they can host bright, short lived {\it bursts} of star formation
which could make these currently dim and inconspicuous galaxies
dominate the luminosity of the Local Group for short periods of time.
Because dwarfs are so numerous, it only requires each galaxy to burst
once or twice in its life time before the dwarfs in the Local group
are effectively always visible in redshift surveys.  This means that
these small galaxies could be dominating redshift surveys in the
intermediate redshift range ({\it e.g.}, Lilly {\it et al.} 1996). The
ubiquitous faint blue galaxies seen in deep imaging and spectroscopic
surveys at intermediate redshifts
could be a population of dwarf galaxies.

\begin{figure} 
\vskip 1cm
\hskip 2.5cm
\psfig{figure=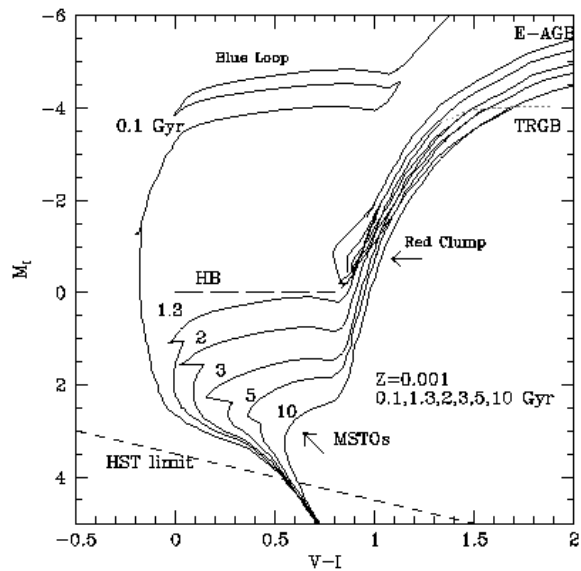,height=8cm,width=8cm}

\caption{ 
Here are plotted theoretical isochrones (from Bertelli {\it et al.} 1994)
for a single metallicity (Z=0.001, or [Fe/H]=$-$1.3) and a range of
ages, as marked in Gyr at the MSTOs to illustrate the CMD features
discussed in \S 3.  Also plotted is the sensitivity limit that HST can
reach for all the galaxies in the Local Group with a modest investment
of time.
}
\end{figure} 

\section{Colour-Magnitude Diagram Analysis: How to Study Dwarf Galaxies}

The study of resolved stellar populations provides a powerful tool to
follow galaxy evolution consistently and directly in terms of physical
parameters such as age (star formation history), chemical
composition and enrichment history, initial mass function (IMF),
environment, and dynamical history of the system.  Photometry of
individual stars in at least two filters and the interpretation of
Colour-Magnitude Diagram (CMD) morphology gives the least ambiguous
and most accurate information about variations in star formation
within a galaxy back to the oldest stars (see Figure~2).  Some of the
physical parameters that affect a CMD are strongly correlated, such as
metallicity and age, since successive generations of star formation
may be progressively enriched in the heavier elements. Careful,
detailed CMD analysis is a proven, uniquely powerful approach ({\it e.g.},
Tosi {\it et al.} 1991; Tolstoy \& Saha 1996; Aparicio {\it et al.}
1996; Mighell 1997; Dohm-Palmer {\it et al.} 1997, 1998; 
Gallagher {\it et al.}  1998; Tolstoy {\it et al.}
1998; Tolstoy 1998a) that benefits enormously from the high spatial
resolution of HST.

\subsection{Useful Features in a Colour-Magnitude Diagram}
Stellar evolution theory provides a number of clear predictions, based
on relatively well understood physics, of features expected in CMDs
for different age and metallicity stellar populations.  There are a
number of clear indicators of varying star formation rates at
different times which can be combined to obtain a very accurate
picture of the entire star formation history of a galaxy (see
Figures~2 and 3). Here I provide a brief description of each of the
separate indicators, in order of preference. The indicators are
thus presented in an order which broadly represents the
ease with which age and metallicity information can be extracted.

\subsubsection{Main Sequence Turnoffs (MSTOs)}
The Main Sequence is a well understood mass-luminosity-lifetime
relation, which allows us to extract (relatively) unambiguous information
about the star formation rate with time over the lifetime of a galaxy.
With exposures (going down to M$_V \sim +4$) of the resolved stellar
populations in nearby galaxies we can obtain the {\it unambiguous age
information that comes from the luminosity of MSTOs} back to the
oldest ages.  The MSTOs do not overlap each other and hence provide
the most direct, accurate information about the star formation history
of a galaxy (see Gallart {\it et al.} 1999).  MSTOs can clearly distinguish
between bursting star formation and quiescent star formation.  The age
resolution that is possible does vary, becoming coarser going back in
time, and can also be affected by metallicity evolution.  Our ability
to disentangle the variations in star formation rate depends upon the
the intensity of the past variations and how long ago they occurred and
which filters are used for observation.

\subsubsection{The Core-Helium Burning Blue Loop Stars (BLs)}
Stars of low metallicity and intermediate mass go on extensive ``Blue
Loop'' (BL) excursions after they ignite He in their core.  Stars in
the BL phase are several magnitudes brighter than when on the main
sequence (M$_V \lsim -1$).  The shape of these ``loops'' are a strong
function of metallicity and age.  They thus provide a more luminous
opportunity to accurately determine the age and metallicity of the
young stellar population (in the range, $\sim$~1~Gyr old) in nearby
low metallicity galaxies The luminosity of a BL star is fixed for a
given age, and thus subsequent generations of BL stars do not over-lie
each other, and can be used to trace {\it spatial} variations in
recent star formation over a galaxy ({\it e.g.}, Dohm-Palmer {\it et
al.} 1997b).  The lower the metallicity of the galaxy, the older will
be the oldest BLs and the further back in time an accurate spatially
resolved star formation history can easily be determined.

\subsubsection{The Red Giant Branch (RGB)}
The RGB is a bright evolved phase of stellar
evolution, where the star is burning H in a shell around its He
core. It is characterized by a fairly constant maximum (or tip)
luminosity (at M$_I = -4.$), and stars are distributed all the way
down to M$_I \sim +2$). Metallicity is the most important effect in
determining the width of the RGB in colour, especially for ages
$>$~2\ Gyr.  However, correlations between age and metallicity can mask
a metallicity spread, as $\sim 4$\ Gyr of age difference can produce the
same effect as 0.1dex of metallicity difference, in (V$-$I).  For a
{\it given} metallicity the RGB blue and red limits are given by the age
spread of the stars populating it (ages $\gsim$1~Gyr), because as a
stellar population ages the RGB moves to the red.  However increasing
the metallicity of a stellar population will also make the RGB redder,
and thus produce the same effect as aging.  This is the (in)famous
age-metallicity degeneracy problem.  The result is that if there is
metallicity evolution within a galaxy, it is impossible to uniquely
disentangle effects due to age and metallicity on the basis of the
optical colours of the RGB alone.

\subsubsection{The Red Clump/Horizontal Branch (RC/HB)}
Red Clump (RC) stars (M$_V \sim +0.5$), low-mass analogues of the BL
stars, and their lower mass cousins the Horizontal Branch (HB) stars
(M$_V \sim 0.$) are core He-burning stars, and they don't obey a
simple mass-luminosity law, as their core mass is mostly independent
of their total mass.  Their luminosity and colour varies depending
upon age, metallicity and mass loss (Caputo, Castellani \&
degl'Innocenti 1995).  The extent in luminosity of the RC can be used
to estimate the age of the population that produced it (see Tolstoy
1998b).  This age measure is {\it independent of absolute magnitude
and hence distance}.

The classical RC and RGB appear in a population at about the same time
(after $\sim$ 0.9--1.5 Gyr, depending on model details), where the RGB
are the progenitors of the RC stars.  The lifetime of a star on the
RGB, t$_{RGB}$, is a strongly decreasing function of M$_{star}$, but
the lifetime in the RC, t$_{RC}$ is roughly constant.  Hence the
ratio, t$_{RC}$ / t$_{RGB}$, is a decreasing function of the age of
the dominant stellar population in a galaxy, and the ratio of the
numbers of stars in the RC, and the HB to the number of RGB is
sensitive to the star formation history of the galaxy ({\it e.g.}
Cole 1999; Gallagher {\it et al.} 1998; Tolstoy {\it et al.} 1998; Han
{\it et al.} 1997).  Thus, the higher the ratio, N(RC)/N(RGB), the
younger the dominant stellar population in a galaxy.

The presence of a large HB population on the other hand (high
N(HB)/N(RGB) or even N(HB)/N(MS)), is caused by a predominantly much
older ($>$10~Gyr) stellar population in a galaxy.  The HB is the
brightest unambiguous indicator of very lowest mass (hence oldest)
stellar populations in a galaxy, it is however impossible to use it to
infer star formation rates at these ancient epochs, because of the
``second parameter effect'' ({\it e.g.}, Fusi Pecci \& Bellazzini
1997), which decouples the HB lifetimes from initial conditions, is
well known, but not yet understood, from globular cluster studies.

\begin{figure} 
\vskip 1cm
\hskip -.6cm
\psfig{figure=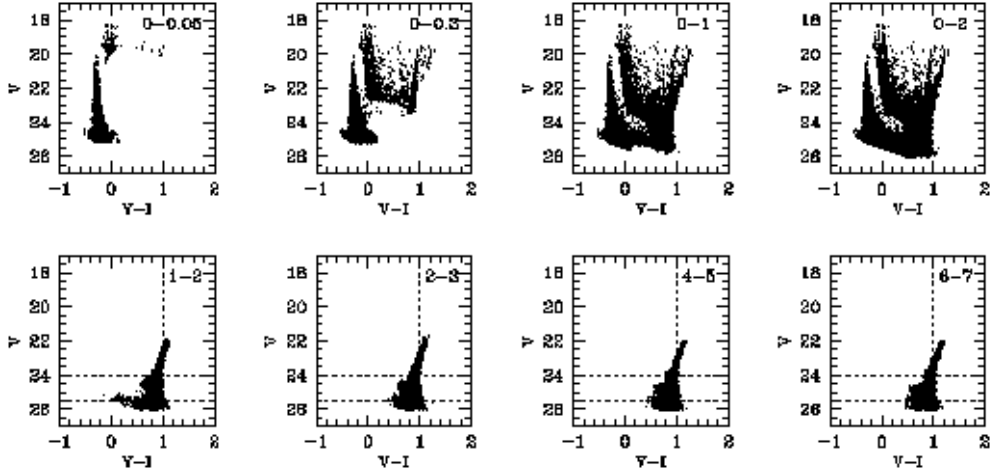,height=6.5cm,width=14cm}
\vskip 1cm
\caption{
Model CMDs, from Monte-Carlo simulations, assuming a standard IMF.
based upon theoretical stellar evolution tracks at metallicity,
Z=0.0004 (or [Fe/H]=~$-$1.7), from Fagotto {\it et al.} (1994).  The periods
of star-formation are marked in the top right-hand corner of each
figure in Gyr. Each period is assumed to have a constant star
formation rate. All the models are made for the sensitivity of WFPC2
over two orbits (one in F555W, one in F814W) to a resolved stellar
population at a distance modulus, (m-M)$_o = 24.2$ (or 700kpc), which is
about the mean for dwarf galaxies in the Local Group.
}
\end{figure} 

\subsubsection{The Extended Asymptotic Giant Branch (EAGB)}

Extended Asymptotic Giant Branch (EAGB) stars are very bright, red
evolved stars (M$_V > -4.$ and typically $V-I > 1.5$).  The
temperature and colour of the EAGB stars in a galaxy are determined by
the age and metallicity of the population they represent.  However
there remain a number of uncertainties in the comparison between the
models and the data ({\it e.g.}, Gallagher et al. 1998; Lynds {\it et
al.} 1998).  It is necessary that more work is done to enable a better
calibration of these very bright indicators of past star formation
events. The future of this field probably lies in infra-red
observations of these stars.

\begin{table} 
  \begin{center} 
  \caption{HST CMDs of Local Group Dwarf Galaxies}

  \begin{tabular}{ll} 
\\
{\it Dwarf Ellipticals:}& \\
	M~32		& Grillmair {\it et al.} 1996 \\	
        NGC~205		& Jones {\it et al.} 1996 \\	
	NGC~185		& Geisler {\it et al.} in prep \\	
	NGC~147 	& Han {\it et al.} 1997 \\	
	Sagittarius	& Mighell {\it et al.} 1997\\	
\\			  
{\it Dwarf Irregulars:} & \\
	NGC~6822	& Wyder {\it et al.} 2000 \\	
	IC~1613		& Cole {\it et al.} 1999; Dolphin {\it et al.} 2000\\	
	IC~5152		& snap-shot in archive\\	
	IC~10		& Hunter {\it et al.} in prep\\
	Sextans~B	& no data\\	
	Sextans~A	& Dohm-Palmer {\it et al.} 1997a,b\\	
       	WLM   		& Dolphin 2000; Rejkuba {\it et al.} 2000\\ 
	Phoenix		& Holtzman {\it et al.} 2000\\	
	Pegasus		& Gallagher {\it et al.} 1998\\	
	LGS~3		& Miller {\it et al.} in prep\\	
	Leo~A		& Tolstoy {\it et al.} 1998\\	
	SagDIG		& no data\\	
	DDO~210		& no data\\	
	EGB 0427+63	& no data\\
\\			  
{\it Dwarf Spheroidals:}& \\
	Fornax		& Buonanno {\it et al.} 1999 \\	
	Ursa~Minor	& Mighell \& Burke 1999; Feltzing {\it et al.} 1999;\\
		 	& Hernandez {\it et al.} 2000\\
	Draco		& Grillmair {\it et al.} 1998\\	
	Leo~I		& Gallart {\it et al.} 1999; Hernandez {\it et al.} 2000 \\
	Sextans		& no data\\	
	Carina		& Mighell 1997; Hernandez {\it et al.} 2000 \\	
	Sculptor	& Monkiewicz {\it et al.} 1999\\	
	Antlia		& no data\\	
	Tucana		& Seitzer {\it et al.} in prep \\	
	Cetus		& no data\\	
	Leo~II		& Mighell \& Rich 1996; Hernandez {\it et al.} 2000 \\	
	And~I		& Da Costa {\it et al.} 1996\\	
	And~II		& Da Costa {\it et al.} 2000\\	
	And~III		& Da Costa {\it et al.} in prep\\	
	And~V		& Armandroff {\it et al.} in prep\\
	And~VI      	& Armandroff {\it et al.} in prep\\
        And~VII		& snap-shot in archive\\

  \end{tabular}
  \end{center} 
\end{table} 

\subsection{Monte-Carlo Simulations of CMDs}

One of the most impressive advances in interpreting observed CMDs in
terms of a detailed star formation history has come from the technique
of re-creating an observed CMD from model stellar evolution tracks by
means of Monte-Carlo simulations. This technique has the advantage
that it can account for the many uncertainties which plague our
understanding of a CMD in what is arguably the most physically
realistic manner.  This approach was pioneered by Tosi and
co-workers in Bologna ({\it e.g.}, Tosi {\it et al.} 1991), and has since been
used and adapted as the standard method of CMD analysis ({\it e.g.}, Tolstoy
\& Saha 1996; Aparicio {\it et al.} 1996; Dolphin 1997; Hernandez et
al. 1999).

The main uncertainties in the interpretation of CMDs of nearby
galaxies come from: estimates of the distance of the galaxy; the
foreground and internal extinction, both the absolute values, and the
patchiness can be uncertain; metallicity, and how this might vary in
time within the galaxy; the initial mass function, what it is and if
it might vary; the fraction of binary stars in the CMD; photometric
errors, or the accuracy with which measurements can be made;
incompleteness, which is a measure of how the number of stars detected
per resolution element affects both the determination of photometric
errors and the number of stars of different luminosities that will be
hidden behind and in the wings of brighter neighbours; and last but
not least and perhaps most difficult of all the uncertainties in the
theoretical models of stellar evolution.  One problem for the modelers
is to find useful data sets to compare with models. Globular clusters
are excellent test data for checking low metallicity and very old
models, and open clusters are mostly quite metal rich.  Dwarf galaxies
tend to be dominated by intermediate and even young metal-poor stellar
populations, and so it is hard to find fiducials.  The result is that
we are forced to try and understand what is happening with an
uncertain star formation history and uncertain model effects all at
the same time.  Any one of the uncertainties listed here could (and
have) produced (long) papers in their own right. They can have 
profound effects on the star formation history determined from a CMD.

\begin{figure} 
\hskip -.55cm
\psfig{figure=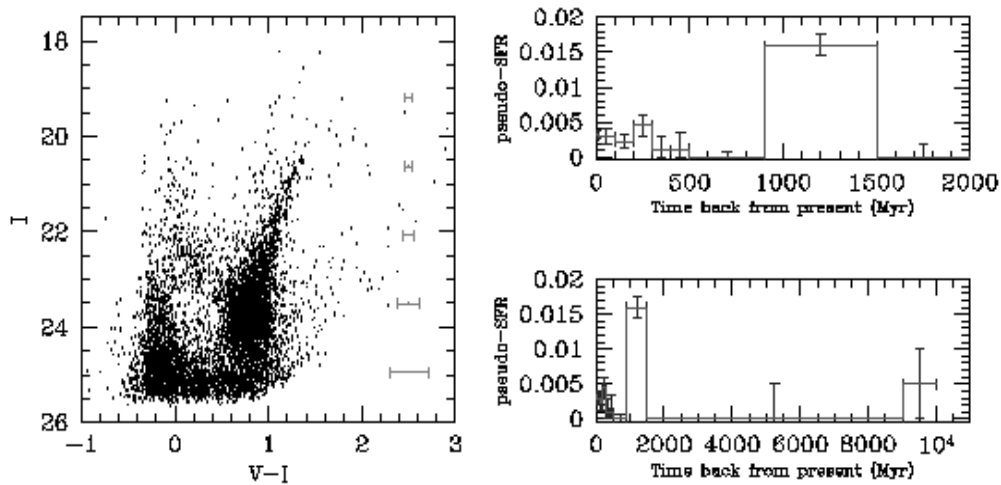,height=7cm,width=14cm}
\caption{
A Leo~A (I, V$-$I) CMD from two orbits of WFPC2 observations (one in
filter F555w and one in F814W), from Tolstoy {\it et al.} (1998). Also shown
is a possible star formation history from detailed modeling of the
CMD, and consistent with the known details of the stellar population,
such as the luminosity spread of the RC; the width of the RGB; and the
presence of BLs. The error bars shown are not in any specific sense
statistical, they merely give an indication of how flexible the star
formation rate at any time can be before significantly affecting the
good match of this model to the data (see Tolstoy {\it et al.} for details).
}
\end{figure} 

So, clearly a great deal of care has to be taken not to over-interpret
CMDs, because having so many factors which affect the modeling means
that there is a lot of parameter space to explore, and the effect of
all the uncertainties is to smear out features which can result in a
very shallow minimum to any $\chi^2$ estimate or Likelihood function
to find the best model.  Each change of initial assumptions requires
generation of a complete new set of models and goodness of fit
assessment. This can be computationally intensive and, in the end, it
can be difficult to find a unique solution. Best solutions are only
ever one out many that are possible. It is thus important to tie in
star formation rate variations to distinct and well determined
features in a CMD, and to justify all variations that are seen.  This
is especially true in older populations where very small and subtle
changes can have dramatic impacts on assumed age and metallicity
variations ({\it c.f.} Tolstoy \& Saha 1996; Tolstoy {\it et al.} 1998),
and see Figure~3.

In Figure~3, I have created a series of model CMDs (see Tolstoy 1996),
which, for a constant metallicity, show the variations that age can
give to a CMD distribution. In the younger populations this is
obvious, but as older populations dominate it takes very accurate
photometry to distinguish between dramatically different ages of
stellar populations.  If metallicity is involved this becomes even
more tricky because age and metallicity can produce very similar
effects on an RGB (see \S 3.1.3).  The CMDs in Figure~3 are made
assuming a galaxy at 700~kpc, and only one orbit of integration time
in each filter. This means there is no MSTOs older than about 800Myr,
which explains why there is such similarity between CMDs which have
differing proportions of population older than about 800~Myr.

Despite this dramatic list of problems, surprisingly detailed and
robust results have come out from the analysis of CMDs. In the next
section I will illustrate some of the most impressive CMDs and their
analyses which have come from HST data for all the different classes of
dwarf galaxy, as defined in \S 1.1.

\begin{figure} 
\psfig{figure=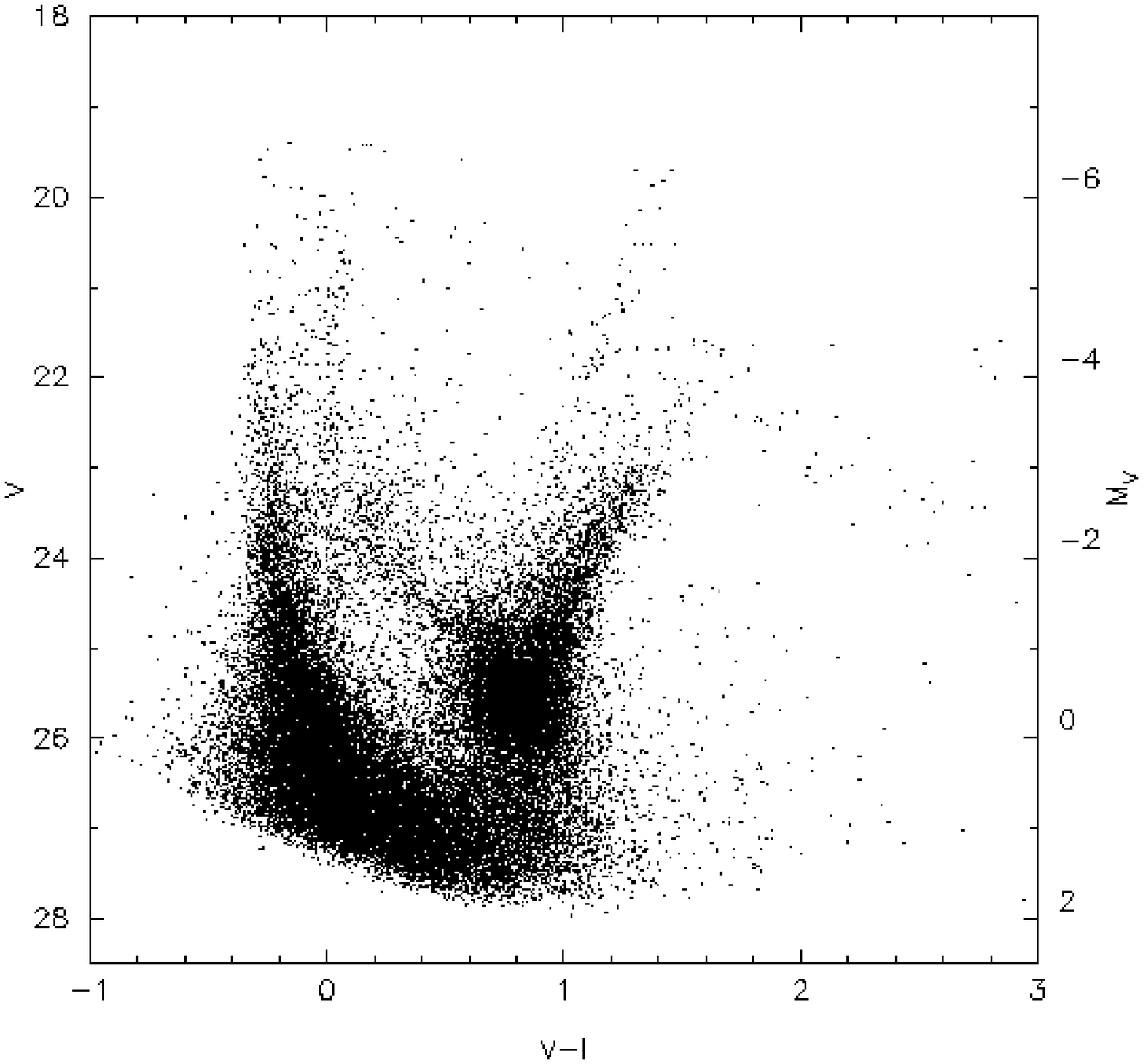,height=6cm,width=6cm}
\vskip -6cm
\hskip 7cm
\psfig{figure=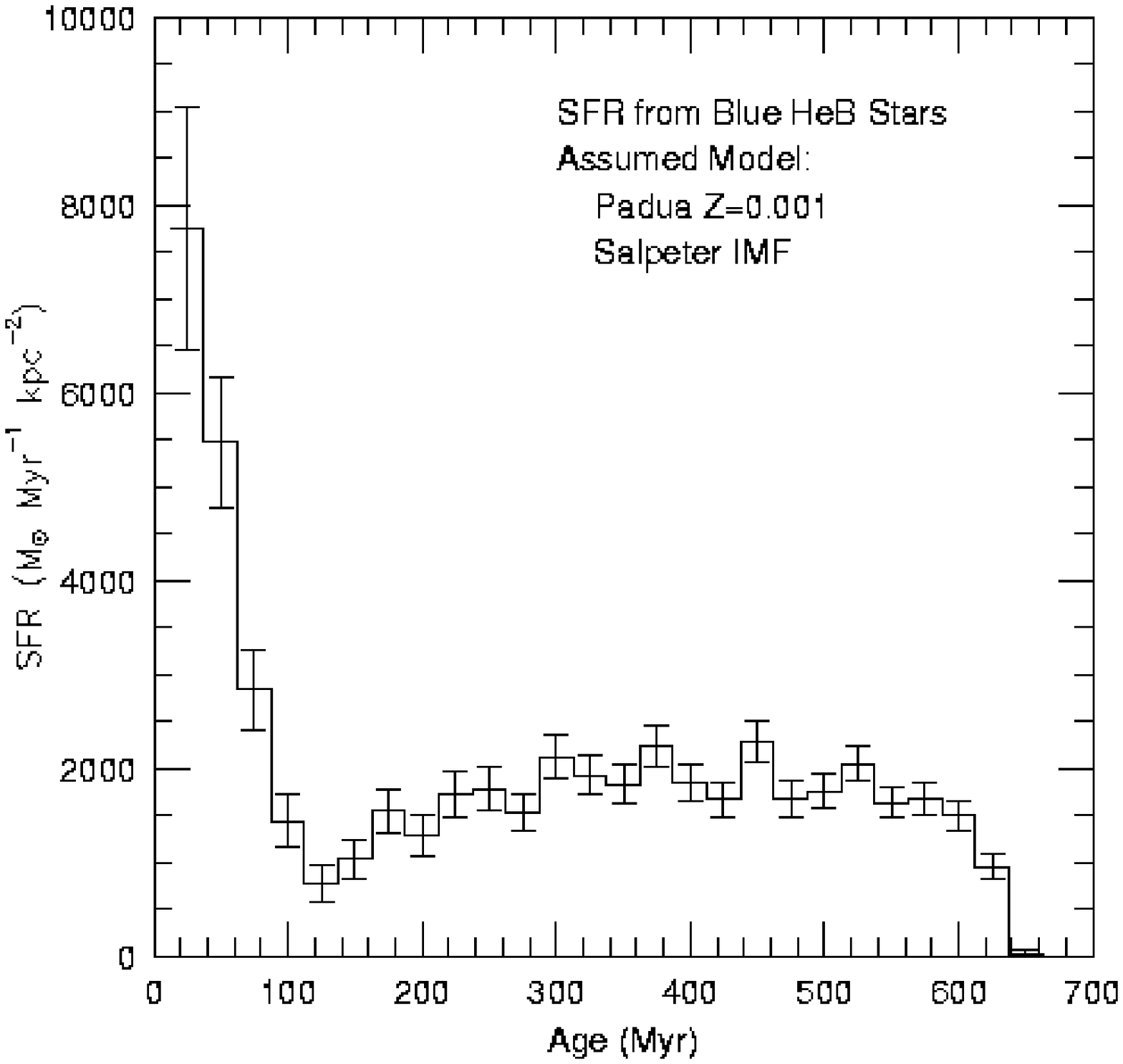,height=6cm,width=6cm}
\caption{
On the right-hand side is the new (V, V$-$I) CMD of Sextans~A from
Dohm-Palmer {\it et al.} 2000, in prep.  It consists of 8 orbits in
F555W and 16 in F814W, and shows the presence of the RC and goes
further down the Main Sequence than the previous 2-orbit CMD of
Dohm-Palmer {\it et al.} (1997). On the left-hand side is the recent
star formation history of Sextans~A determined from the BL stars 
by Dohm-Palmer {\it et al.} (1997).
 }
\end{figure} 

\begin{figure} 
\psfig{figure=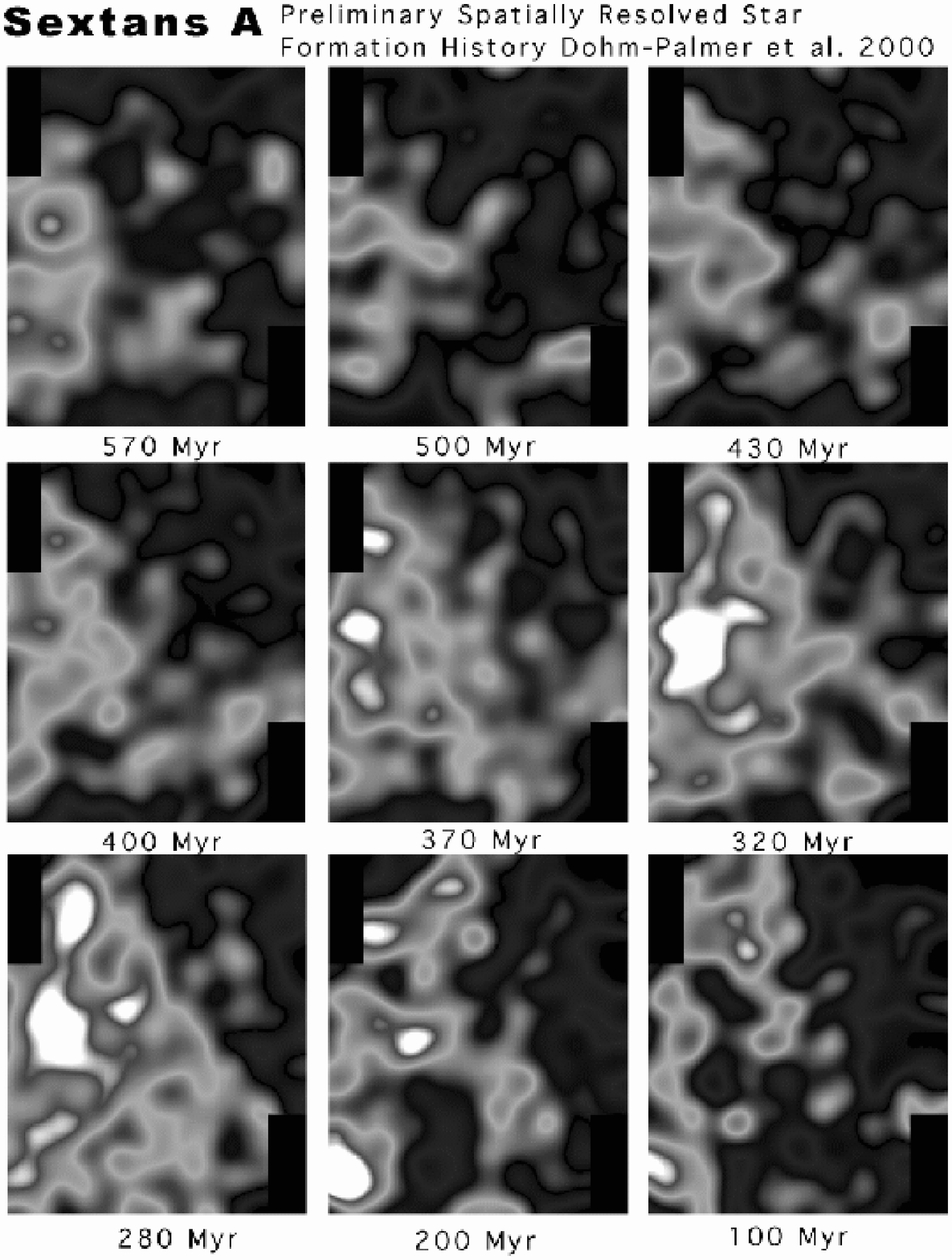,height=17cm,width=13.5cm}
\caption{
Here we see a gallery of movie still frames created by Robbie
Dohm-Palmer (c.f. Dohm-Palmer {\it et al.} 1997), which show how the
intensity of star formation varies spatially with time across
Sextans~A.  These are preliminary results, which combine the the two
WFPC2 pointings on Sextans~A so that the spatial coverage of nearly
the whole luminous centre of the galaxy is achieved.  The time
separation between each frame is not uniform, and was chosen to
high-light interesting features. See Dohm-Palmer {\it et al.} 1997;
2000 for more details. The newer data is on top, and despite the
presence of young blue stars and H-$\alpha$, this does not seem to be
representative of a global increased star formation rate in this part
of the galaxy, unlike the lower area which has vigorous star formation
over the last several hundred million years. 
}
\end{figure} 

\section{HST Observations of Local Group Dwarf Galaxies}

There have been a number of spectacular results from HST imaging of
the resolved stellar populations in Local Group dwarf galaxies (see
Table~2).  In the next sections I provide examples of some of the most
beautiful HST CMDs for a selection of dwarf galaxies in the Local
Group, and just beyond:

\subsection{Leo~A}
Leo~A (DDO~69) is a gas-rich dI galaxy, with an extremely low HII
region abundance ($\sim$3\% solar, van Zee, Skillman \& Haynes 1999).
The interpretation of a CMD from two orbits of WFPC2 data (Tolstoy
{\it et al.} 1998; see Figure~4) is based upon extremely low
metallicity (Z=0.0004; or [Fe/H]$= - 1.7$) theoretical stellar
evolution models (Fagotto {\it et al.} 1994; see also Figure~3), which
suggest that this galaxy is predominantly young, {\it i.e.}, $<$~2~Gyr
old.  A major episode of star formation 900$-$1500~Gyr ago can explain
the RC luminosity and also fits in with the interpretation of the
number of anomalous Cepheid variable stars seen in this galaxy.  The
presence of an older, underlying globular cluster age stellar
population could not be ruled out with these data, however, using the
currently available stellar evolution models, it would appear that
such an older population is limited to no more than 10\% of the total
star formation to have occurred in the centre of this galaxy.
Theoretical models of the chemical evolution of dwarf galaxies by
Ferrara \& Tolstoy (2000) imply that, even though this galaxy is
extremely metal-poor, an underlying older stellar population is
required to build up the current metallicity.  Perhaps this older
population resides in an outer halo. Of course, neither the chemical
evolution models nor the existing CMDs can distinguish between an old
population which formed in a large burst, or more sedate and roughly
constant rate through-out a longer time.

\subsection{Sextans~A}
Sextans A (DDO 75) is a gas-rich dI galaxy with a low metal abundance
([Fe/H]$\sim -1.4$), and active star formation, which is located on
the periphery of the Local Group (1.4~Mpc away).  The initial WFPC2
CMD of Sextans~A, based on two orbits of telescope time, shows several
clearly separated populations that align well with stellar evolution
model predictions for a low metallicity system (see Dohm-Palmer et
al. 1997a,b).  This was the first time a BL sequence had been so
definitively identified in a CMD (see the CMD in Figure~5).  
The star formation history from the
main sequence and BL stars (in Figure~5) was determined by Dohm-Palmer
{\it et al.}  for the last 600~Myr using theoretical stellar evolution
models.  The spatial distribution of the BL stars was then used to
determine the spatial variation of the star formation across Sextans~A
with time (see Figure~6).  Figure~6 is a preliminary result including
new WFPC2 data which covers the whole galaxy (Dohm-Palmer {\it et al.} 2000,
in prep).  The modeling concludes that in the past 50 Myr, Sextans A
has had an average star formation rate that is $\sim10$ times that of
the average over the history of the galaxy.  This current activity is
highly concentrated in a young region in the South-East roughly 25 pc
across.  This coincides with the brightest HII regions and the highest
column density of HI.  Between the ages of 100 and 600 Myr ago, the
star formation has been roughly constant at slightly above the average
value.  There are regions (200$-$300pc across) with a factor of $\sim
5$ enhancement in star formation rate with a duration of 100$-$200
Myr.

\begin{figure} 
\hskip -.6cm
\psfig{figure=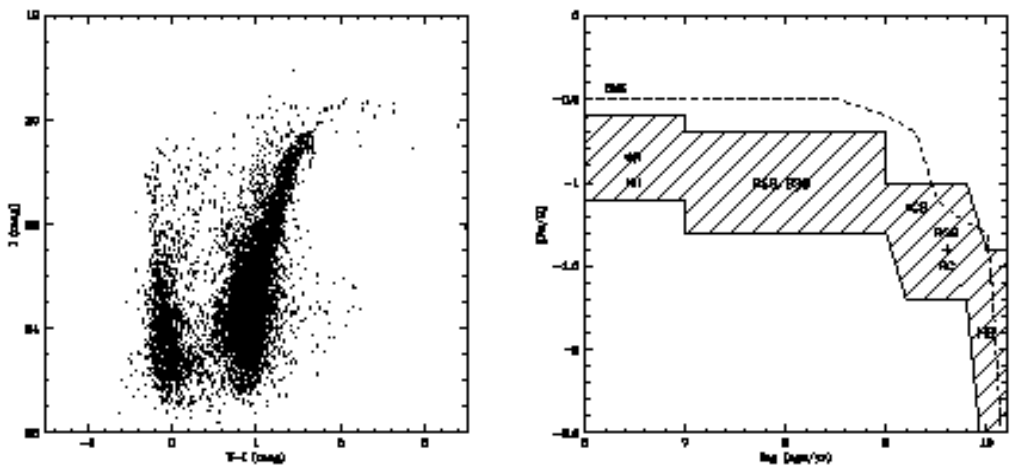,height=7cm,width=15cm}
\caption{
The IC~1613 (I, V$-$I) CMD from 8 orbits of WFPC2 observations (four
in filter F555w and four in F814W) of a field right in the centre of
IC1613, away from any bright HII regions, from Cole et
al. (1999). This was a fairly crowded field and fairly brutal cuts in
S/N were made for this CMD.  Also shown is a possible age-metallicity
relation for IC~1613, also from Cole {\it et al.} Each region of the figure
is labelled with the CMD feature that constrains the metallicity to
lie within the shaded region. Abbreviations are given in the text; and
in addition WR denotes the Wolf-Rayet star; HII denotes the H~II
regions; RSG and BSG are red and blue super giants. The dotted line
shows the mean age-metallicity relation for the SMC.
}
\end{figure} 

\begin{figure} 
\vskip 0.2cm
\hskip 2.7cm
\psfig{figure=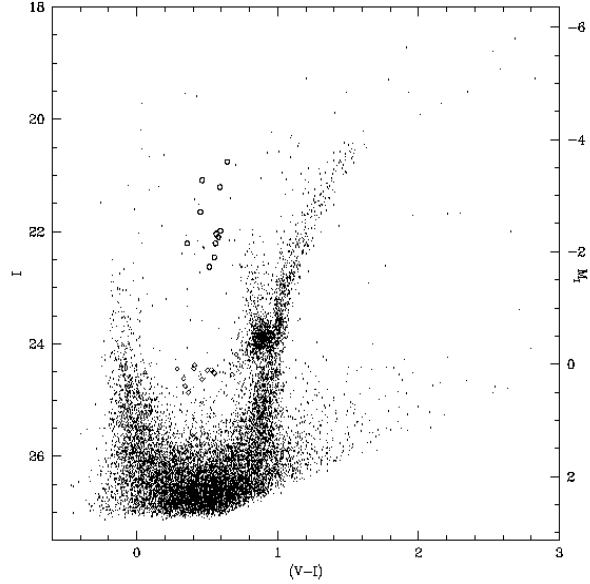,height=8cm,width=8cm}
\caption{
The WFPC2 (I, V$-$I) CMD of the outer field of IC~1613, showing the variable
stars found there (from Dolphin {\it et al.} 2000), coming from 24 orbits, 
16 in F814W and 8 in F555W. The circles represent the
Cepheids, the diamonds the RR Lyraes, and the triangles the two
possible eclipsing binaries.
}
\end{figure} 

\subsection{IC~1613}

IC~1613 is a Magellanic type dI galaxy, with young stars of SMC-like
metallicity ($\sim$10\% solar), which was first resolved into stars by
Baade (1928), it was later used by Baade (1963) to illustrate the
archetypal ``Baade's sheet'' of underlying RGB (population II) stars,
which have now been found to exist in most, if not all Local Group
dwarfs ({\it e.g.}, Sandage 1971; Hodge 1986; Saha 1995).  IC~1613 is to
date the only Local Group dI (excluding the Magellanic Clouds), in
which RR Lyrae variables have been detected (Saha {\it et al.} 1992; 
Dolphin {\it et al.} 2000).

This spatially extended galaxy has had two pointings with WFPC2, one
in the central region (Cole {\it et al.} 1999, see Figure~7) and one in the
outskirts (Dolphin {\it et al.} 2000, Figure~8).  The two CMDs, despite
being in quite different environments within the galaxy, look
remarkably similar.

The main-sequence luminosity function provides evidence for a roughly
constant star formation rate of ($\sim 3.5 \times 10^{-4} \Msun
yr^{-1}$ across the central WFPC2 field of view (0.22 kpc$^2$)) during
the past $\sim$250-350 Myr, and going back to $\sim$1~Gyr in the outer
field (Tolstoy {\it et al.} in prep.).  Structure in the BL function
implies that the star formation rate was $\sim$50\% higher 400-900 Myr
ago than today.  The blue HB was also detected in both IC~1613
pointings, again showing that the ancient stars in this galaxy are
uniformly distributed in the inner and outer regions of IC~1613.  It
was already known that IC~1613 contains a population of RR~Lyrae
variable stars (Saha {\it et al.} 1992) in its outer halo, and hence
an ancient stellar population, but this is the first time that a deep
enough CMD was made to detect the HB.  From the different populations
identified and modeled, an approximate age-metallicity relation for IC
1613 was determined (see Figure~7), which appears, like the present
day metallicity, to be similar to that of the SMC.

The natural HST orbital cadence of 90 minutes is ideal for the
detection of short period variable stars, such as RR~Lyrae.  As the
second WFPC2 pointing in the outskirts of IC~1613 consisted of 8
orbits in F555W and 16 orbits in F814W this data set contains just
enough distinct observations to be able to identify and classify short
period variables in the WFPC2 field of view (Dolphin {\it et al.}
2000).  These are plotted on top of the deeper CMD of the outer region
in Figure~8. There are 13 RR~Lyraes, which unambiguously mark out the
presence of the modestly populated HB, and also 11 short-period
Cepheids, which are indicators of an intermediate-age population in
IC~1613.

\subsection{WLM}
WLM (Wolf-Lundmark-Melote; DDO 221) 
is another large Local Group Magellanic dI galaxy,
with approximately SMC-like present-day metallicity.
It is the only dI galaxy in the Local Group with
a bona-fide globular cluster (see Hodge {\it et al.} 1999).

\begin{figure} 
\vskip 0.2cm
\hskip 0.5cm
\psfig{figure=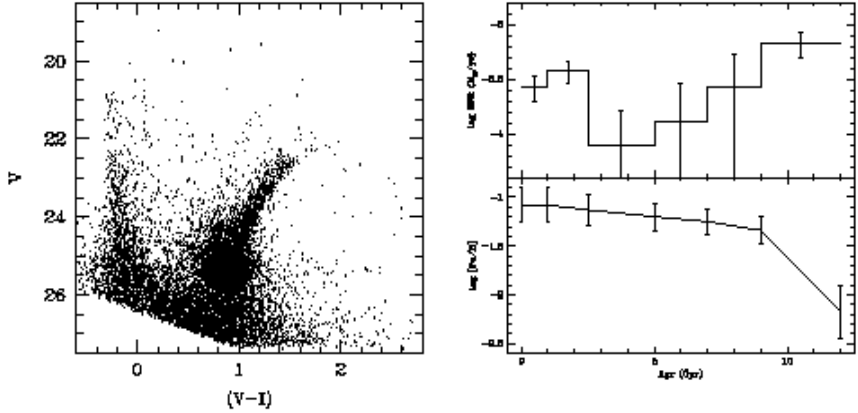,height=6cm,width=12cm}
\caption{
The WFPC2 (V, V$-$I) CMD of WLM, the combined results from photometry
on the 3 WF chips.  The data were taken over 4 orbits, 2 in both F814W
and F555W, from Dolphin (2000).  Also shown are the star formation
history and the chemical evolution history that Dolphin derives from
these data.
}
\end{figure} 

\begin{figure} 
\hskip 1cm
\psfig{figure=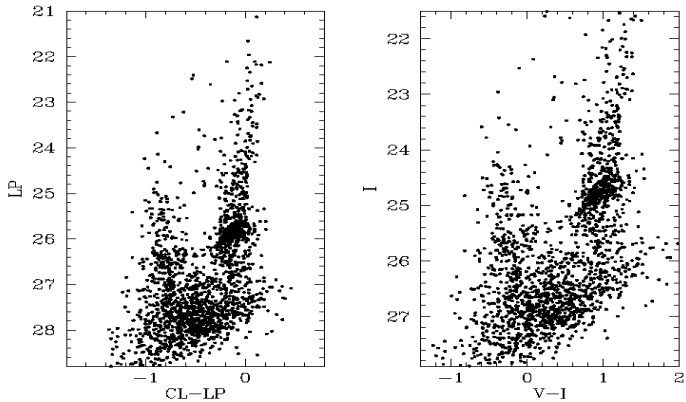,height=6cm,width=10cm,angle=-90}
\caption{
Here are the STIS/CCD imaging results for WLM from Rejkuba {\it et
al.} (2000).  On the left is the CMD in the raw STIS photometric system
(LP, CL$-$LP) and on the right is the CMD after conversion to a
standard photometric system (I, V$-$I).  These data were taken in
parallel to the WFPC2 data shown in Figure~9.
}
\end{figure} 

The HST pointing on WLM in September 1998 must rate as an extremely
efficient one. The WFPC2 PC chip contained the globular cluster, and
was used for a study of this unique object (Hodge {\it et al.} 1999). The
remaining 3 WF chips contained stars from the field population of WLM,
and were used by Dolphin (2000) to study the star formation history of
the field population of this galaxy. The RC is clearly detected (see
Figure~9), it is less clear if there is an HB present. The star
formation history and the corresponding metal enrichment history which
Dolphin determined from the WFPC2 CMD are also shown in Figure~9.  The
models are most reliable for the young and intermediate-age
populations.  Finally, as part of a targeted STIS parallel survey
program, STIS images were also taken in parallel to the WFPC2 primary
observations in both broad band filters available (Clear [CL] and Long
Pass [LP], see Gardner {\it et al.} 1998), at a position about 4$^{\prime}$
away from the WFPC2 position.  The STIS CCD in combination with these
extremely broad filters means the images go fainter than the WFPC2,
however the field of view is also a lot smaller
($\sim25^{\prime\prime} \times 50^{\prime\prime}$).  It had been shown
theoretically that these (very broad) STIS filters can be usefully
transformed to an effective V and I (Gregg \& Minniti 1997), and this
has been confirmed by the results of Rejkuba {\it et al.} 2000 (see Figure
10) using these parallel data from WLM.  They clearly detect a RC
(as Dolphin 2000, in the inner WFPC2 field of view does), but unlike
Dolphin they clearly detect a HB. This is due to the
increased sensitivity of the STIS CCD; the Dolphin WFPC2 CMD barely reaches
the HB magnitude, whereas the STIS CMD clearly goes beyond.

\subsection{Leo~I}
Leo~I is a nearby (250~kpc) relatively isolated dSph. It has been
poorly studied from the ground because of the proximity of the 1st
magnitude star, Regulus.  However WFPC2 has produced one of its most
detailed CMDs from Leo I (Gallart {\it et al.} 1999; see Figure~11). The CMD
goes down to an absolute magnitude, M$_V = +4.5$ on the Main Sequence.
This depth of CMD allows a very accurate star formation history to be
determined from MSTOs (Gallart {\it et al.} 1999, see Figure~11).
They found that most (70$-$80\%) of star formation
activity in Leo~I occurred between 1 and 7 Gyr ago. A fairly uniform
star formation rate dropped dramatically about a Gyr ago, and around
300Myr it seems to have stopped altogether.  It is not clear from
these results whether or not Leo~I contains an ancient ($>$10~Gyr old)
stellar population. It does not have an obvious HB, and it is one of
the few dSph not to have a detected RR~Lyrae population but a very
large population of anomalous Cepheids (Lee {\it et al.} 1993). This
is consistent with the Gallart {\it et al.} models which find that this
galaxy is dominated by an intermediate-age, metal poor, stellar population.

\begin{figure} 
\vskip 2.cm
\psfig{figure=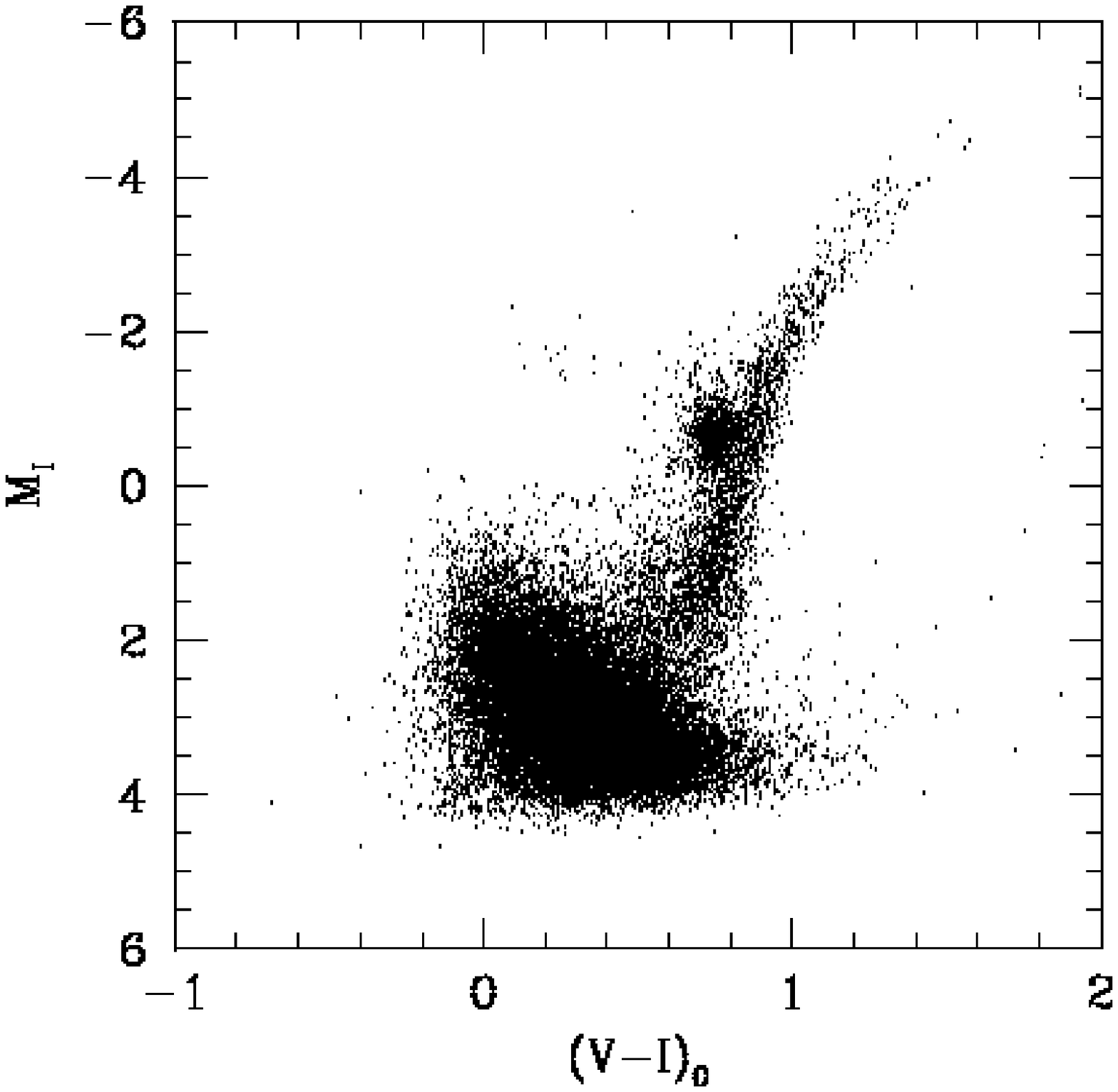,height=7cm,width=7cm}
\vskip -8.8cm
\hskip 6.8cm
\psfig{figure=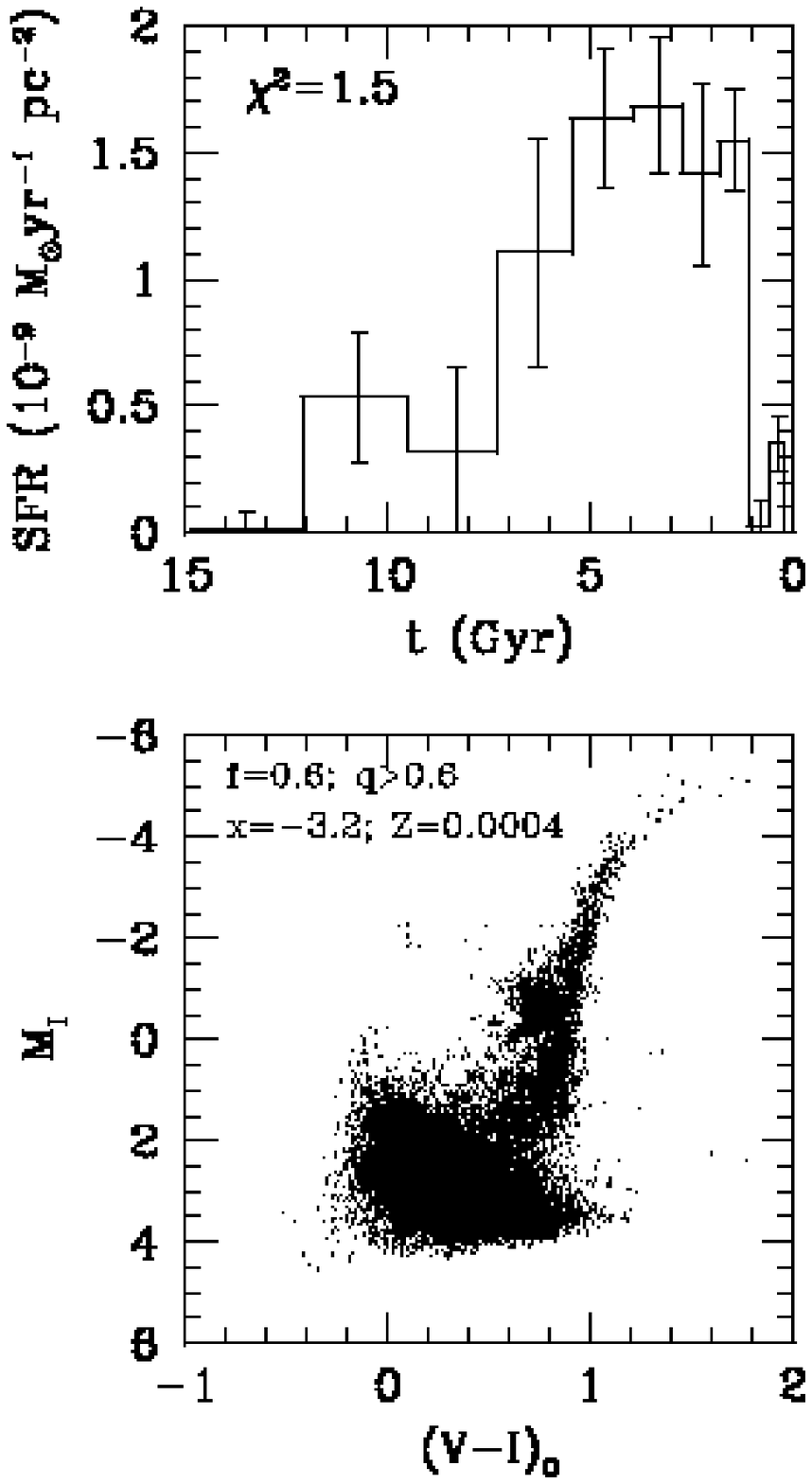,height=9.8cm,width=6.5cm}
\caption{
The WFPC2 (I, V$-$I) CMD of Leo~I is shown here, assuming a distance
modulus of (m-M)=22.18, from Gallart {\it et al.} (1999).  These data come
from 6 orbits of exposure time, 3 in both F555W and F814W.  Also shown
is the star formation history that Gallart {\it et al.} found to be the best
match to the data, and the resulting model CMD form this history.
They come from a model with metallicity Z$=$0.0004 ([Fe/H]$= -1.7$),
an IMF slope, x$= -3.2$, and a carefully determined binary fraction,
see Gallart {\it et al.} for more details.
}
\end{figure} 
%
\begin{figure} 
\vskip 0.2cm
\hskip 0.5cm
\psfig{figure=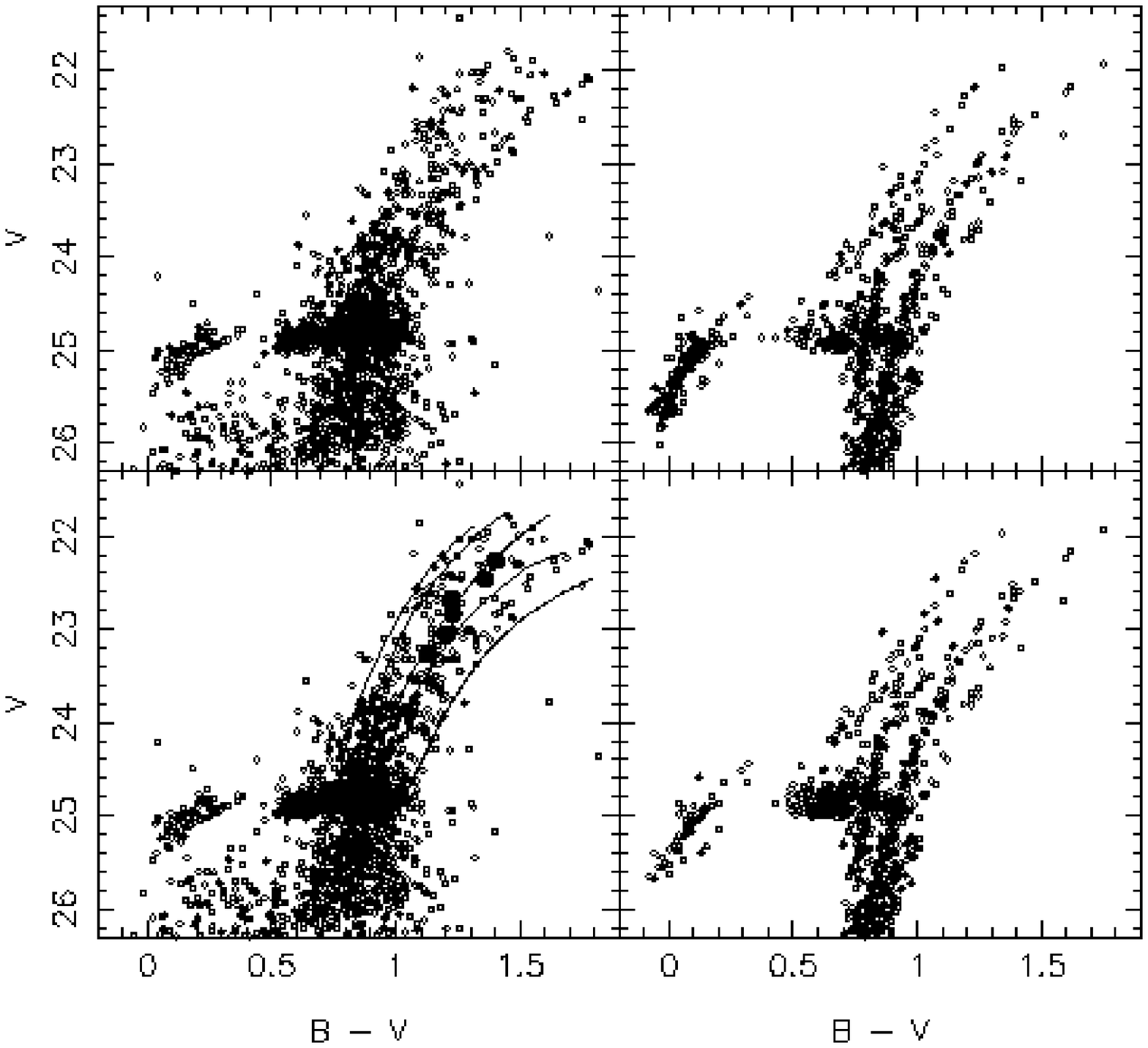,height=12cm,width=12cm}
\caption{
Here are presented, on the left-hand side, the WFPC2 CMDs for 11
orbits of exposure time on Andromeda~II transformed to the standard
(V, B$-$V) magnitude system, from Da Costa {\it et al.} (2000). The
candidate variables have been excluded from these plots.  The lower
left-hand CMD is identical to the upper, but shown superposed with
the giant branches of the standard globular clusters M68 ([Fe/H] $=
-$2.09), M55 ([Fe/H] $= -$1.82), NGC 6752 ([Fe/H] $= -$1.54), NGC 362
([Fe/H] $= -$1.28), and 47 Tuc ([Fe/H] $= -$0.71). The filled symbols
give the mean And~II RGB colours in $\pm$0.1 V magnitude bins.  On the
right-hand side, the upper panel is a composite CMD made up from
observed CMDs of the Galactic globular clusters M~55 ([Fe/H] $=
-$1.82), NGC~1851 ([Fe/H] $= -$1.29), and 47~Tuc ([Fe/H] $= -$0.71),
with the relative star numbers (44/45/11), scaled to reflect the
And~II abundance distribution. This CMD clearly has relatively more
blue HB stars than And~II has. In the lower right-hand panel all the
NGC~1851 blue HB and 40\% of the M~55 blue HB stars have been replaced
with red HB stars from NGC~362 (80\%) and 47~Tuc (20\%), and the HB
morphology of the ``model'' CMD is now a better match to that which is
observed.  See Da Costa {\it et al.} for many more details.
}
\end{figure} 

\subsection{Andromeda~II}
Andromeda~II (And~II) is a dSph companion to M~31.  The WFPC2 CMD (Da
Costa {\it et al.} 2000; see Figure 12) shows a predominantly red HB (like
in most other dSph). In And~II there is no evidence for a radial
gradient, unlike, And~I (Da Costa {\it et al.} 1996), or NGC~147 (Han {\it et
al.} 1997).  In And~II Da Costa {\it et al.}  also detect probable RR~Lyrae
variable stars, although the number of images taken of And~II are not
sufficient to accurately classify the variables found, but those
identified with the colours of the HB can safely be assumed to be
RR~Lyrae variables.  To interpret
the And~II CMD in terms of a star formation history Da Costa et
al. have used a combination of standard Galactic globular cluster CMDs
scaled to reproduce the And II mean abundance and abundance
dispersion, to interpret the observed HB morphology (see Figure~12).
They find that at least 50\% of the total stellar population must be
younger than the age of the globular clusters. This inference is
strengthened by the small number of upper-AGB carbon stars, and the
relatively faint luminosities (M$_{bol}\sim -$4.1) of these stars.
These upper-AGB carbon stars have assumed ages of around 6$-$9 Gyr,
whilst the existence of blue HB and RR Lyrae variable stars argues for
the presence of an old ($>$10 Gyr) population. Thus, And II must
have had an extended epoch of star formation like many of the Galactic
dSphs. The RGB colors yield a mean abundance of
$<$[Fe/H]$>=-$1.49$\pm$0.11 and a surprisingly large internal
abundance spread, of about 0.36~dex.  It is not possible to model the
abundance distribution in And II with single component simple chemical
enrichment model. However, a simple model with a dominant
``metal-poor'' ([Fe/H]$= -$1.6) and a ``metal-rich'' ([Fe/H]$= -$0.95)
component appears to produce the best match (see Figure~12).

\begin{figure}
\vskip 0.2cm
\psfig{figure=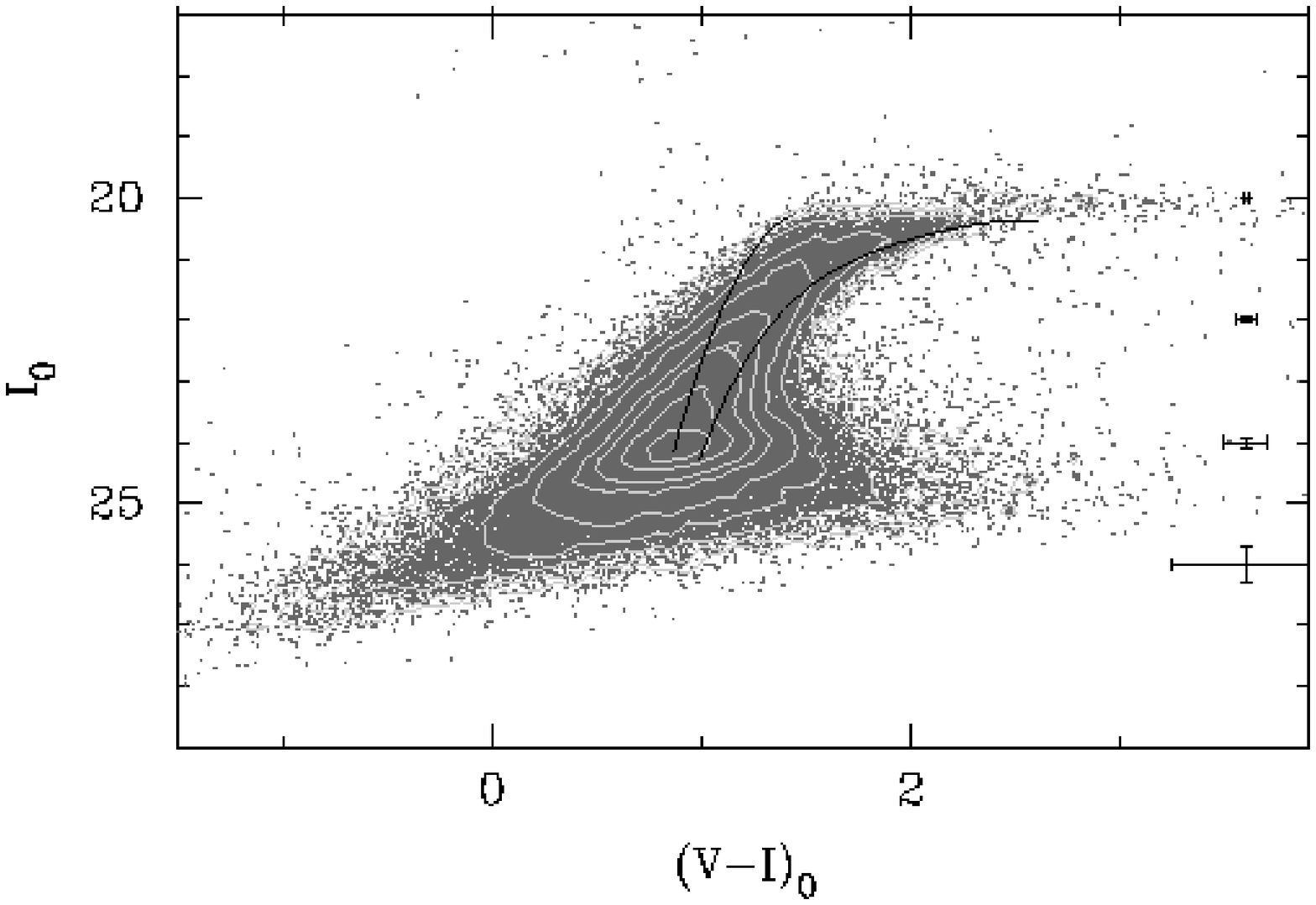,height=5.5cm,width=6.5cm}
\vskip -5.55cm
\hskip 6.5cm
\psfig{figure=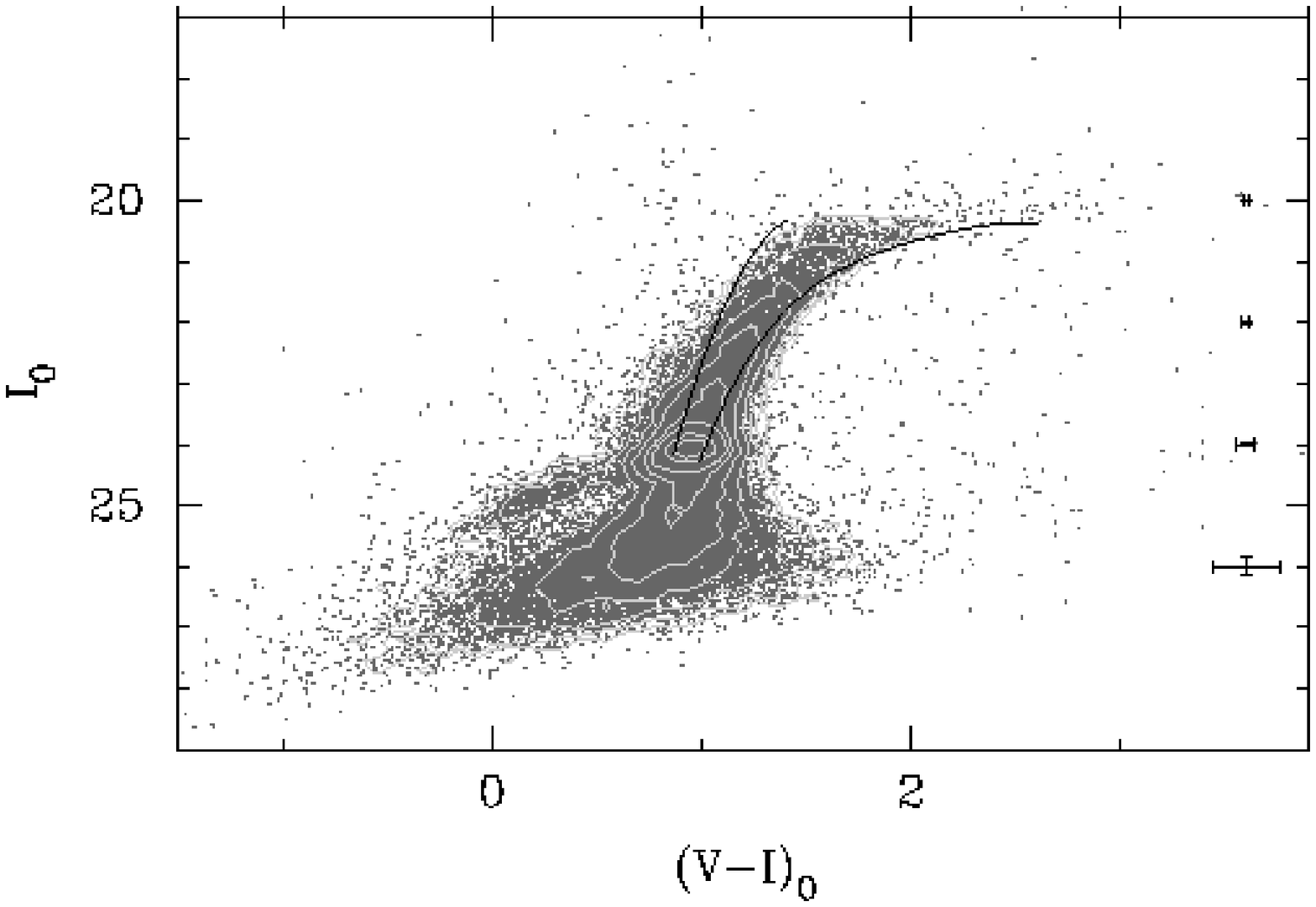,height=5.5cm,width=6.5cm}
\caption{
Here are shown the WFPC2 (I,V$-$I) CMDs for two fields in NGC~147,
from Han {\it et al.} (1997), which come from 9~orbits of telescope time.
On the left-hand side is the CMD for the inner most field of NGC~147,
which does very close into the extremely crowded heart of this small
galaxy. This CMD contains nearly 80 000 photometered stars. This means
that the relative numbers of stars is also indicated by density
contours overlying the points. Also over-plotted are the average error
bars on the photometry at given magnitudes, and also Galactic globular
cluster fiducials for M~15 and 47~Tuc.  This is then done identically
on the right-hand side for the outer field of NGC~147, which contains
only about half the number of stars in the central field. It also
contains a much more distinctive HB, which is not only a function of
the reduced crowding in the outer field. See Han {\it et al.}  for more
details.
}
\end{figure} 

\subsection{NGC~147}
NGC~147 is a dE galaxy associated with M~31.  There are WFPC2 CMDs at
two positions in this galaxy, at inner and outer positions (Han et
al. 1997, see Figure~13). There are significant differences between
the inner and outer field stellar populations (as can be seen at a
glance of Figure~13), and these cannot be explained by differences in
crowding properties of these fields, even though the inner field is
extremely crowded. The RGB suggests a metallicity of [Fe/H]=-0.9 in
the inner, central field and [Fe/H]$= -$1.0 in the outer one, and the
outer field shows a weak tendency of increasing metallicity with
galactocentric radius. The RGB also shows evidence of a metallicity
dispersion in NGC~147, with a larger dispersion closer to the centre
of the galaxy. The age of most of the stars in the RGB is assumed to
be $>$5~Gyr.  The small population of EAGB stars does show the
presence of an intermediate-age population (a few Gyr old), which
seems to be larger towards the centre of the galaxy, contrary to the
bulk of the older stars. The HB stars are more populous towards the
outer part of the galaxy. Again, consistent with an age (and
metallicity) gradient within NGC~147.  The absence of any main
sequence stars shows that any star formation completely ceased at
least a Gyr ago.

\subsection{VII~Zw403}
VII~Zw403 (UGC~6456) is definitely not in the Local Group, but at a
distance of $\sim$4.5Mpc, it is is most likely an isolated member at
the far side of the M~81 group. The WFPC2 CMD (see Figure~14, from
Lynds {\it et al.}  1998) is the best example of a resolved BCD.  Also
shown in Figure~14, is a possible star formation history determined
from a quantitative analysis of the CMD from Lynds {\it et al.}  Another
study of the same HST observations of VII~Zw 403 is presented by
Schulte-Ladbeck {\it et al.} (1999a), and they get similar results. The HST
CMD of this relatively distant galaxy is directly comparable to to
ground based observations of closer dwarf galaxies.  The similarity
between Figure~14 and the ground-based CMD of NGC~6822 of Gallart {\it
{\it et al.}} (1994) is quite startling, especially in the properties of the
EAGB.  Clearly this is a similarity which needs further study.
NGC~6822 is close enough to calibrate the EAGB versus star formation
history determined from the detection of older MSTOs.

VII~Zw403 is also one of the first galaxies to have a NICMOS, IR CMD
(Schulte-Ladbeck {\it et al.} 1999b), which detects a large number of
red super-giant and AGB stars, and reaches the tip of the RGB in J and
H.  In principle extending the colour baseline out from the optical to
the IR offers advantages for separating out different stellar phases
from one another ({\it e.g.} Bertelli {\it et al.} 1994).  However,
without high S/N IR data the photometric errors significantly limit
the improvement.

\begin{figure} 
\psfig{figure=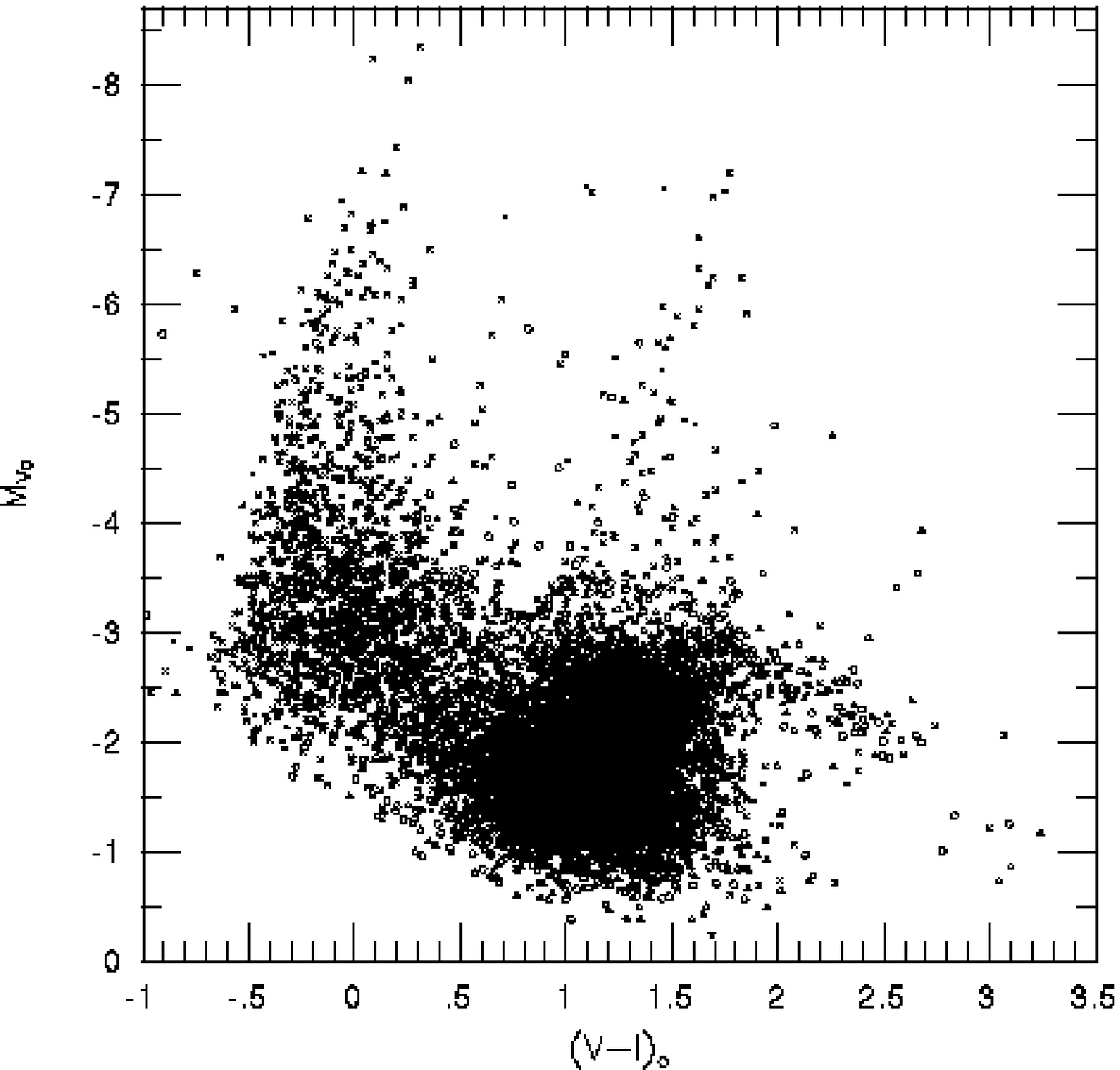,height=6cm,width=6.8cm}
\vskip -6.1cm
\hskip 7.3cm
\psfig{figure=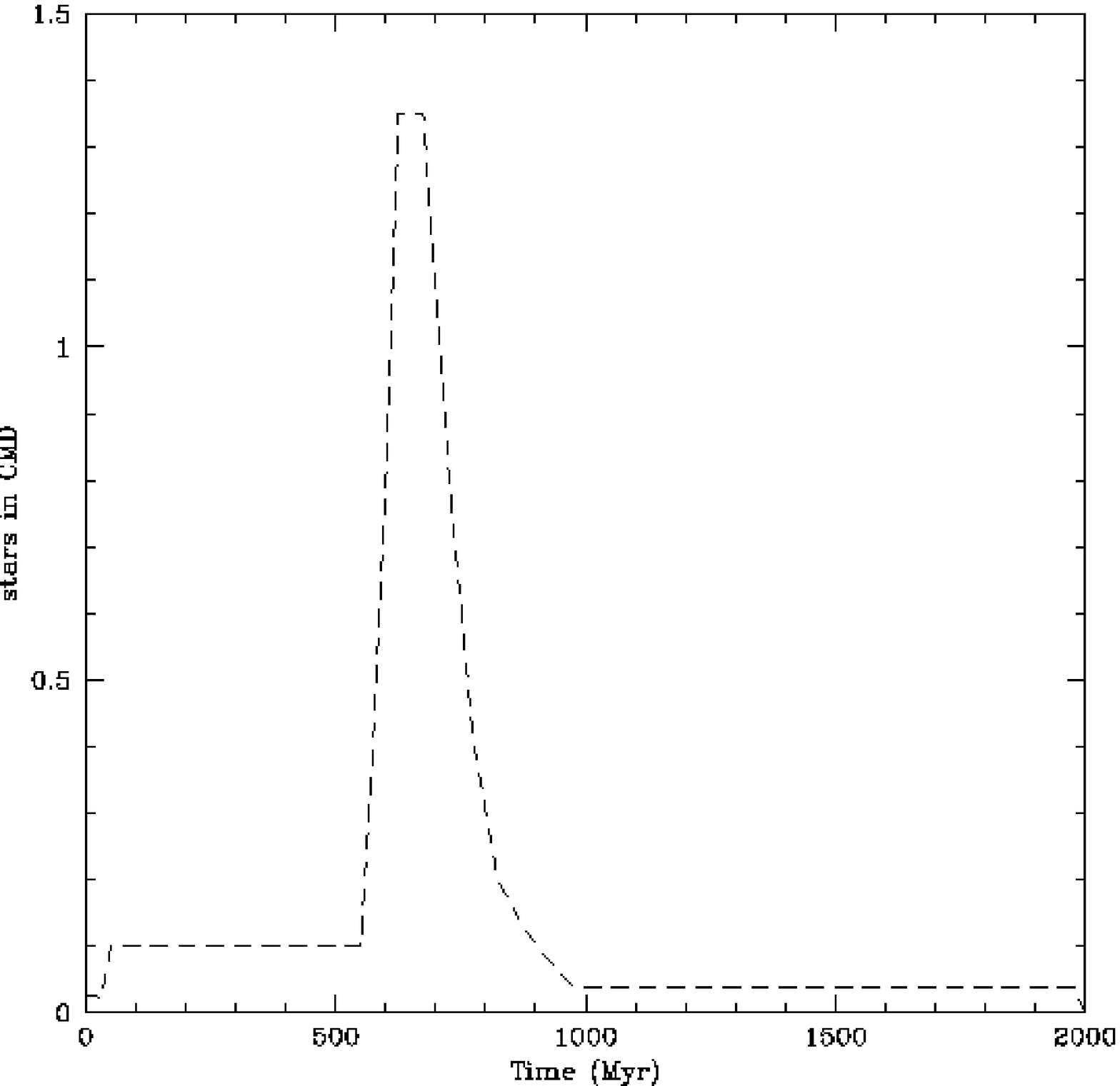,height=5.9cm,width=6.cm}
\vskip0.5cm
\caption{
Here is the WFPC2 (V,V$-$I) CMD for the BCD VII~Zw403, taken from
Lynds {\it et al.} (1998). This CMD is based on four orbits of telescope
time (2 per filter), and also shown is a possible star formation
history based on a careful modeling of the CMD. See Lynds {\it et al.} for
more details.
}
\end{figure} 
%
\begin{figure} 
\vskip 0.7cm
\psfig{figure=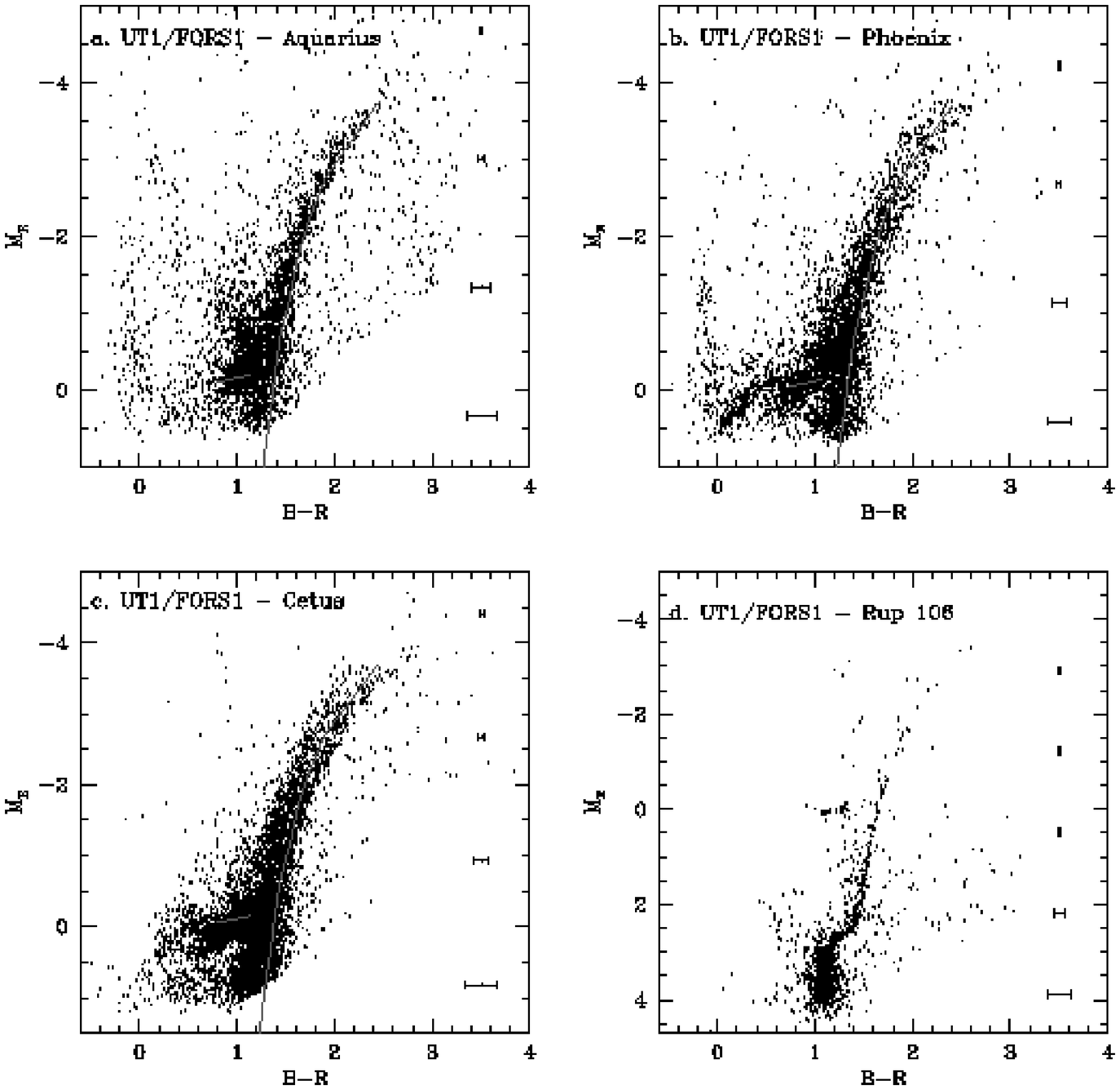,height=13cm,width=13cm}
\caption{ 
Here are four VLT/UT1 (R, B$-$R) CMDs for three Local Group galaxies
(DDO~210, Cetus and Phoenix) and one youngish (12$-$13~Gyr old)
globular cluster (Ruprecht~106), from Tolstoy {\it et al.} (2000). DDO~210
and Cetus had exposure times of 3000sec in R and 3600sec in B; roughly
equivalent to a total of 3 orbits of HST. Phoenix received only
1600sec in R and 1800sec in B, and Ruprecht~106 30sec in R and 80sec
in B. Representative error bars are plotted for each data set, and
the fiducial RGB and HB from Ruprecht~106 data are over-plotted on each
of the galaxy CMDs.
}
\end{figure}

\section{New Results from a New Telescope}

In the future, the large, ground based telescopes can play a vital
complementary role to the HST. With their relatively wide 
fields of view and 
larger apertures, ground based telescopes will become the
instruments of choice for imaging the extended and less crowded
halo populations of the nearby galaxies. Large telescopes on
the ground are also ideal for spectroscopic
follow-up on individual stars in a CMD to determine abundances
and the internal dynamics of nearby galaxies.
To show that competition from new ground based facilities is coming
along fast, in Figure~15 I show CMDs for three Local Group galaxies
(and a calibration globular cluster) from the VLT, and the FORS1
imaging/spectrograph. These data were taken in excellent seeing
conditions in August 1999, and have been published in preliminary form
by Tolstoy {\it et al.} (2000).  The galaxies Cetus, Aquarius
(DDO~210), and Phoenix were selected because they are relatively
nearby, open-structured dI/dSph systems.  Because of the excellent
seeing conditions we were able to obtain very deep exposures covering
the densest central regions of these galaxies, without our images
becoming prohibitively crowded.  From these images we have made very
accurate CMDs of the resolved stellar population down below the
magnitude of the HB region.  In this way we have made the first
detection of RC and/or HB populations in these galaxies, which reveal
the presence of intermediate and old stellar populations.  In the case
of Phoenix we detect a distinct and populous blue HB, which indicates
the presence of quite a number of stars $>$10~Gyr old.  These results
further strengthen evidence that most, if not all, galaxies no matter
how small or metal poor contain some old stars.  Another striking
feature of our results is the marked difference between the
Colour-Magnitude diagrams of each galaxy, despite the apparent
similarity of their global morphologies, luminosities and
metallicities.  For the purposes of accurately interpreting our
results we have also made observations in the same filters of a
Galactic globular cluster, Ruprecht~106, which has a metallicity
similar to the mean of the observed dwarf galaxies.

\section{Conclusions}

A survey of the resolved stellar populations of all the galaxies in
our Local Group provides a uniform picture of the global star
formation properties of galaxies with a wide variety of mass,
metallicity, gas content etc., and makes a sample that ought to
reflect the star formation history of the Universe and give results
which can be compared to high redshift survey results ({\it e.g.}, Steidel
{\it et al.} 1999). Initial comparisons suggest these different
approaches do not yield the same results (Tolstoy 1998b; Fukugita {\it
et al.} 1998), but the errors are large due to the lack of {\it
detailed} star formation histories of nearby galaxies.  The CMDs
presented here whilst beautiful and dramatic represent the tip of the
iceberg for the Local Group. Most of the observations reported here
consisted of 1 or 2 orbits of integration time per filter.  To really
complete a detailed census of the nearby resolved stellar populations
we need to go as deep as the sensitivity limit given in Figure~2 for
all Local Group galaxies, to detect the oldest MSTOs. This includes
the large Local Group galaxies, such as M~31, which along with our
Galaxy represent the dominant mode of star formation in the Local
Group. With data like this we will know the star formation history of
the Local Group, going back to the earliest times.  We have also shown
that data from ground based telescopes in excellent seeing with active
optics can compete with HST images, and indeed make an excellent
complement, by being more blue sensitive, and having a larger field of
view. It is only with the deepest exposures of the most crowded
regions that HST is still the undisputed
winner, 
because it is hard to gain enough exposure time in excellent
stable seeing conditions on the ground, as
no where under the Earth's atmosphere
can ideal conditions be guaranteed.  
It looks promising that Adaptive Optics on large
ground based telescopes will one day rival HST supremacy, 
but this is an endeavour that will be restricted to
infra-red wavelengths for the foreseeable future.
HST must lead the way to extend
detailed star formation studies past the Galaxy and its immediate
satellites. 


\begin{acknowledgments}
{\it Acknowledgments:} 
I would like to thank those who generously provided me with postscript
files of their figures, to my specifications, to use in this article,
namely: Gary Da Costa; Carme Gallart; Marina Rejkuba; Andrew Cole;
Robbie Dohm-Palmer; Deidre Hunter; Andy Dolphin and Jay Gallagher.  I
would also like to thank Evan Skillman, Jay Gallagher and Andrew Cole
for useful discussions about many of the details of form and content
presented here.  Last but not least I would like to thank the
organisers of the HST Symposium for inviting me; it was an honour and
a pleasure to put together and present this talk.
\end{acknowledgments}

\end{document}